\documentclass[useAMS,usenatbib,usegraphicx,usedcolumn]{mn2e}

\usepackage{fixltx2e}

\newcommand{\CIV}{\mbox{C\,{\sc iv}}}

\newcommand{\SVI}{\mbox{S\,{\sc vi}}}

\newcommand{\SiIV}{\mbox{Si\,{\sc iv}}}
\newcommand{\AlIII}{\mbox{Al\,{\sc iii}}}

\newcommand{\NV}{\mbox{N\,{\sc v}}}

\newcommand{\OVI}{\mbox{O\,{\sc vi}}}

\newcommand{\MgII}{\mbox{Mg\,{\sc ii}}}

\newcommand{\PV}{\mbox{P\,{\sc v}}}
\newcommand{\HeII}{\mbox{He\,{\sc ii}}}

\newcommand{\kms}{km~s$^{-1}$}

\newcommand{\ergs}{erg~s$^{-1}$}
\newcommand{\cmmt}{cm$^{-2}$}
\newcommand{\cmt}{cm$^{-3}$}
\newcommand{\LLedd}{\mbox{$L/L_{\rm Edd}$}}
\newcommand{\Lion}{\mbox{$L_{\rm ion}$}}

\newcommand{\aox}{\mbox{$\alpha_{\rm ox}$}}
\newcommand{\aion}{\mbox{$\alpha_{\rm ion}$}}

\newcommand{\NH}{\mbox{$\Sigma_{\rm H}$}}
\newcommand{\NHtot}{\mbox{$\Sigma_{\rm H}$}}
\newcommand{\Nion}{\mbox{$N_{\rm ion}$}}

\newcommand{\hnu}{\mbox{$\langle h\nu\rangle$}}

\newcommand{\rdust}{\mbox{$r_{\rm dust}$}}

\newcommand{\dd}{\textrm{d}}
\newcommand{\cl}{{\sc cloudy}}

\title[RPC -- IV. Application to BALs]{Radiation pressure confinement -- IV. Application to broad absorption line outflows}

\author[A.~Baskin, A.~Laor and J.~Stern]
{
Alexei Baskin,$^1$\thanks{E-mail: alexei@physics.technion.ac.il} 
Ari Laor$^1$ and Jonathan Stern$^2$\\
$^1$ Physics Department, Technion -- Israel Institute of Technology, Haifa~32000, Israel\\
$^2$ Max-Planck-Institut f\"{u}r Astronomie, K\"{o}nigstuhl 17, D-69117 Heidelberg, Germany
}

\begin{document}
\date{}
\pagerange{\pageref{firstpage}--\pageref{lastpage}} \pubyear{2014}
\maketitle
\label{firstpage}

\begin{abstract}
A fraction of quasars present broad absorption lines, produced by outflowing gas with typical velocities of 3000--10,000~\kms. If the outflowing gas fills a significant fraction of the volume where it resides, then it will be highly ionized by the quasar due to its low density, and will not produce the observed UV absorption. The suggestion that the outflow is shielded from the ionizing radiation was excluded by recent observations. The remaining solution is a dense outflow with a  filling factor $f<10^{-3}$. What produces such a small $f$? Here we point out that radiation pressure confinement (RPC) inevitably leads to gas compression and the formation of dense thin gas sheets/filaments, with a large gradient in density and ionization along the line of sight. The total column of ionized dustless gas is a few times $10^{22}$~\cmmt, consistent with the observed X-ray absorption and detectable \PV\ absorption. The predicted maximal columns of various ions show a small dependence on the system parameters, and can be used to test the validity of RPC as a solution for the overionization problem. The ionization structure of the outflow implies that if the outflow is radiatively driven, then broad absorption line quasars should have $\LLedd\ga0.1$.
\end{abstract}

\begin{keywords}
galaxies: active -- quasars: absorption lines -- quasars: general.
\end{keywords}

\section{Introduction}\label{sec:intro}

Broad absorption line quasars (BALQs) display broad blueshifted absorption of resonance lines, with outflow velocities often reaching $\sim $30,000~\kms\ \citep*{weymann_etal81, reichard_etal03, trump_etal06, gibson_etal09}, with a few cases extending up to $\sim$50,000~\kms\  \citep*{hamann_etal97, rodriguez_hidalgo_etal11}. Only in a few rare cases there is broad redshifted absorption, generally of a few 1000~\kms\ \citep{hall_etal13}. The absorption profile is sometimes broad and smooth, extending over a velocity range of $>$10,000~\kms\ \citep{turnshek_etal88}. However, more commonly the absorption trough is composed of a number of distinct troughs, each only a few 1000~\kms\ broad (e.g.\ \citealt{korista_etal92, hamann_98, arav_etal01b, gabel_etal06, trump_etal06, gibson_etal09}). 

The outflow generally resides outside the Broad Line Region (BLR), as it absorbs both the continuum and the broad emission lines (cf.\ \citealt{arav_etal99}). Based on variability it does not likely reside much further outside the BLR \citep{capellupo_etal11, capellupo_etal12, filiz_ak_etal13}. This is in contrast with associated narrow absorption lines, with width $<$1000~\kms, which are likely associated with gas much further out (e.g.\ \citealt{de_kool_etal02, moe_etal09, edmonds_etal11}). However, one cannot exclude a large radius for some BAL outflows (e.g.\ \citealt{borguet_etal13}). 

The wind driving mechanism may either be radiation pressure on lines (e.g.\ \citealt*{arav_etal94, murray_etal95, proga_etal00}), radiation pressure on dust \citep*{voit_etal93, scoville_norman95}, MHD outflows (e.g.\ \citealt{konigl_kartje94, everett_05, fukumura_etal10}), or a combination of e.g.\ magnetic and radiation pressure \citep{de_kool_begelman95}. However, a persistent problem in all wind models, regardless of the driving mechanism, is the overionization problem. This results from the following simple argument.
The ionization state of the wind is set by the ionization parameter 
\begin{equation}
U\equiv \frac{n_{\gamma}}{n_e},
\end{equation}
where $n_{\gamma}$ is the ionizing photon density, and $n_e$ is the electron density. About half of the bolometric luminosity ($L_{\rm bol}$) of AGN is ionizing, with a mean energy $\hnu\simeq 2.5$~Ryd, which implies 
\begin{equation}
n_{\gamma}=2.6\times 10^7L_{46}r_{\rm pc}^{-2} \mbox{~\cmt}, \label{eq:int_n_gamma}
\end{equation}
where $L_{\rm bol}=10^{46}L_{46}$ and $r_{\rm pc}$ is the distance from the ionizing source in pc. If the absorber extends over a distance comparable to the distance from the continuum source, then a uniform-density absorber with a H column density of $\NH=10^{22}\Sigma_{\rm H, 22}$~cm$^{-2}$, has a density of 
\begin{equation}
n_e=3200 \Sigma_{\rm H, 22} r_{\rm pc}^{-1} \mbox{~\cmt}.
\end{equation}
The absorber ionization parameter is then 
\begin{equation}
U=8.1\times 10^3L_{46}r_{\rm pc}^{-1}\Sigma_{\rm H, 22}^{-1}.\label{eq:intro_U}
\end{equation}
Prominent absorption is observed for example from the C$^{3+}$ ion, which requires $U<1$ to be detectable (e.g.\ \citealt{hamann_97}).  Thus, even for a large column absorber with  $\NH=10^{23}$~cm$^{-2}$, an extended uniform-density absorber will be too highly ionized to produce \CIV\ absorption, out to $r\sim 1$~kpc. 

\citet{murray_etal95} suggested that the overionization problem is avoided by foreground gas which filters the ionizing radiation. Such a filter needs to suppress the ionizing continuum by a factor of $>10^3$ to suppress $U$ (eq.~\ref{eq:intro_U}) enough to avoid overionization (see also \citealt{chelouche_netzer03}). BALQs do tend to show weaker X-ray emission \citep*{brandt_etal00}, but generally by a factor of only 10--30. Recent high energy X-ray observations suggest the X-ray weakness is intrinsic and not due to absorption \citep{teng_etal14}. Furthermore, the recent study of \citet{hamann_etal13} of mini-BALs with extreme velocity outflows, finds only weak X-ray absorption, which is not sufficient to prevent overionization. These observations exclude the radiative shield mechanism as a possible solution for the overionization problem. Current hydrodynamical models of BAL winds also fail to produce a radiative shield which prevents overionization \citep{higginbottom_etal14}. The only remaining solution is a highly clumped wind, with a low enough filling factor of $f<10^{-3}$, which increases $n_e$ and lowers $U$ enough to prevent overionization.

In addition, broad absorption lines display a wide range of ionization levels, which excludes a single uniform-density absorber \citep{turnshek_etal96, hamann_97}. Given the observed similar kinematics of the different ions, the different absorbers likely have a similar spatial distribution, which lead to the suggestion of a multiphase outflow \citep*{everett_etal02}.

Below we show that radiation pressure compression (RPC) of the outflowing gas, naturally explains the highly clumped nature of BAL outflows. RPC leads to absorbers in the form of thin `pancakes' viewed face-on, as envisioned by \citet{hamann_etal13}. RPC also produces a radial density profile in the absorber, which leads to a range in $U$ in a given absorber system. Such a range in $U$ at a given position can be misinterpreted as a multiphase outflow. However, in a multiphase medium one expects the different phases to be in pressure equilibrium, while in RPC the pressure increases with decreasing $U$ (see Section~\ref{sec:model_RPC}). We also point out that the ionization structure of the absorber is effectively independent of distance, and a similar mechanism may be relevant for absorbers on significantly larger scales.

In Section~\ref{sec:model} below, we briefly review RPC, estimate the expected absorbing column by this mechanism, and describe the numerical calculation. The numerical results are presented in Section~\ref{sec:results} and discussed in Section~\ref{sec:discussion}. A brief summary is given in Section~\ref{sec:conclusions}.

\section{Theory}\label{sec:model}

\subsection{Radiation Pressure Compression}\label{sec:model_RPC}
The RPC mechanism was first applied by \citet{dopita_etal02} to the narrow line region (NLR) in AGN. In paper~I \citep*{stern_etal14}, we applied this method to gas emission on pc to kpc scale, and in paper~II \citep*{baskin_etal14} to the gas in the BLR. In paper III \citep{stern_etal14b} RPC is applied to X-ray warm absorbers, and in this paper we apply it to UV absorbers observed in BALQs. RPC is described in detail in papers I and II. We present here a brief description of the mechanism, and discuss its applicability to BAL outflows. Note that in papers I and II we used RPC as an acronym for Radiation Pressure Confinement. Here we term the same mechanism as Radiation Pressure Compression, to stress the compression of the outflowing gas, which is the essence of this paper. The term confinement was used earlier to stress the confinement mechanism of the non-outflowing BLR and NLR gas.

RPC assumes a non accelerating slab of photoionized gas, i.e.\ gas in hydrostatic equilibrium. Gravity is cancelled by the centrifugal force in the rotating frame of the gas, and thus the only external force acting on the gas is the radiation force. The radiation force at a given depth in the gas is balanced by the local gas pressure gradient,
\begin{equation}
 \frac{\dd P_{\rm gas}(r)}{\dd r}=\frac{L_{\rm ion}}{4\pi r^2 c}e^{-\tau(r)}n\bar{\sigma},  \label{eq:dP_dr}
\end{equation}
where  \Lion\ is the ionizing luminosity\footnote{The relevant luminosity might be somewhat larger than \Lion, e.g.\ by a factor of $\simeq$2 for dust absorption, if non-ionizing photons are also attenuated.}, $r$ is the distance from the continuum source, $n$ is the gas density, $\bar{\sigma}$ is the flux-weighted attenuation cross section per particle, $c$ is the speed of light and $\tau(r)$ is the flux-weighted optical depth from the face of the cloud located at $r_{\rm s}$, i.e.\
\begin{equation}
 \tau(r)=\frac{\int \tau_\nu(r)\, L_{\rm ion,\nu}\,e^{-\tau_\nu(r)}\dd\nu} {\int L_{\rm ion,\nu}\,e^{-\tau_\nu(r)}\dd\nu},
\end{equation}
where
\begin{equation}
 \tau_\nu(r)=\int^r_{r_{\rm s}}  n\sigma_\nu \dd r,
\end{equation}
\begin{equation}
\sigma_\nu \equiv \sigma_\nu^{\rm abs}+\sigma_\nu^{\rm sct},
\end{equation}
and $\sigma_\nu^{\rm abs}$  and $\sigma_\nu^{\rm sct}$ are the frequency dependent absorption and electron-scattering cross section per particle, respectively. The gas density and temperature structure can be solved for, utilizing equation~\ref{eq:dP_dr} and the energy equation, as derived from photoionization modelling. The gas pressure of the slab increases inwards, and reaches a maximal value of 
\begin{equation}
P_{\rm gas}=\frac{F_{\rm rad}}{c}+P_{\rm s}
\end{equation}
at $\tau \gg 1$, where 
\begin{equation}
 F_{\rm rad}=\frac{L_{\rm ion}}{4\pi r_{\rm s}^2}
\end{equation}
is the flux incident on the surface of a slab of gas at a distance $r_{\rm s}$, and $P_{\rm s}$ is the pressure at the surface due to the ambient medium. If the incident radiation pressure satisfies
\begin{equation}
P_{\rm rad}=\frac{F_{\rm rad}}{c}\gg P_{\rm s}, 
\end{equation}
then the cloud structure is set by $P_{\rm rad}$, and is independent of $P_{\rm s}$. 

Since $P_{\rm gas}=2n_e kT$ and  $P_{\rm rad}=\hnu n_{\gamma}$, we generally get that
\begin{equation}
U=\frac{P_{\rm rad}}{P_{\rm gas}}\times \frac{2kT}{\hnu}. 
\end{equation}
Deep enough in the slab, where all the incident flux is absorbed [$\tau(r)\gg 1$], the gas pressure inevitably builds up to $P_{\rm gas}=P_{\rm rad}$. Since photoionized gas near the H ionization front is at $T\sim 10^4$~K, RPC leads to a universal $U\simeq 0.1$ in this region, independent of distance from the continuum source. 

Closer to the surface of the slab, in regions where $\tau(r)< 1$, the fraction of radiation absorbed is $\tau(r)$, and therefore 
\begin{equation}
P_{\rm gas}=P_{\rm rad}\tau(r) ,
\end{equation}
which gives
\begin{equation}
U=\frac{2kT}{\tau(r)\hnu}, \label{eq:U_vs_tau}
\end{equation}
or
\begin{equation}
U(r)\simeq 0.1\frac{T_4}{\tau(r)}, \label{eq:U_vs_tau_T}
\end{equation}
where $T=10^4T_4$~K is the gas temperature. For example, at $\tau(r)=10^{-3}$, the RPC photoionization solution gives $T\simeq 10^6$~K (paper I, fig.~2 there), which implies $U=10^4$, and the gas is fully ionized. At $\tau(r)=0.01$, the RPC solution gives $U=10^3$, and the gas is very highly but not fully ionized. At $\tau(r)=0.1$, $T\simeq 10^5$~K,  $U=10$ and the gas is rather highly ionized. From this layer inwards we find the ions which can produce the observed UV absorption lines. The structure of $U(\tau)$ is nearly independent of distance from the ionizing source.

The ionization structure can be viewed as a superposition of uniform-density optically-thin slabs, starting at a high $U$ at the surface of the slab, given by the boundary value of $P_{\rm s}$, and ending with an optically thick slab at $U\simeq 0.1$. Thus, a \emph{single} RPC slab produces a broad range of ionization states (e.g.\ from Mg$^{9+}$ to C$^+$), in contrast with a uniform-density slab, where $U$ is constant, and only deep enough where $\tau(r)\gg 1$ the ionization level drops significantly. 

Thus, RPC provides a natural solution for the overionization problem. The incident radiation pressure compresses the gas, leading to $P_{\rm gas}\simeq P_{\rm rad}$ deep enough, and thus to $U\simeq 0.1$. The observed broad absorption lines originate from gas at $U<10$, or $n_e>0.1n_{\gamma}$. Using the above relation for $n_{\gamma}$ (eq.~\ref{eq:int_n_gamma}) we get  
\begin{equation}
n_e>2.6\times 10^6L_{46}r_{\rm pc}^{-2} \mbox{~\cmt}. \label{eq:ne_for_U10}
\end{equation}
The value of $n_e$ implies a relative thickness of the absorbing layer, $D=\NHtot/n$, of
\begin{equation}
\frac{D}{r}<1.25\times 10^{-3} \Sigma_{\rm H,22} L_{46}^{-1}r_{\rm pc} .
\end{equation}
Thus, the absorbing gas forms thin sheets in the radial direction. However, if the absorbing gas resides on kpc scale, then $D/r\sim 1$, and the gas does not need to be compressed in order to have
$U<10$, i.e.\ avoid overionization.

\subsubsection{Applicability of RPC to BAL outflows}

The hydrostatic RPC solution is valid for a non accelerating outflow. Since the gas is subject to an external force of $F_{\rm rad}/c$ by the incident radiation, there must be a counter acting external radial force directed inwards, to balance this force. A plausible mechanism is ram pressure on the outer surface of the outflowing gas by a low density ambient gas. The ram pressure acting on the leading edge of the absorbing outflow is 
\begin{equation}
 P_{\rm ram}\simeq m_{\rm p} n_{\rm s} v^2,
\end{equation}
where $m_{\rm p}$ is the proton mass, $n_{\rm s}$ is the number density of the ambient gas and $v$ is the outflow velocity. The gas pressure deep inside the slab is
\begin{equation}
 P_{\rm gas}= \frac{F_{\rm rad}}{c}.
\end{equation}
In order for $P_{\rm ram}$ to balance $P_{\rm gas}$, $n_{\rm s}$ should be
\begin{equation}
 n_{\rm s} \approx 10^4 \left(\frac{r}{\rdust}\right)^{-2} \left(\frac{v}{10^4\mbox{~\kms}}\right)^{-2} \mbox{~\cmt}, \label{eq:n_amb}
\end{equation}
where
\begin{equation}
 \rdust=0.2L_{46}^{0.5}\mbox{~pc}
\end{equation}
is the dust sublimation radius \citep{laor_draine93}. The BAL outflow $r$ was estimated for only a handful of objects, and the lowest values are $r\sim 10 \rdust$ \citep{moe_etal09, hamann_etal13}. Substituting this value into equation~\ref{eq:n_amb} implies that an ambient gas with $n_{\rm s} \sim 10^2$~\cmt\ can produce $P_{\rm ram}$ which will balance the absorbed $P_{\rm rad}$ and create a hydrostatic RPC BAL outflow. The low value of $n_{\rm s}$ implies $U\sim 10^5$, and as a result ambient gas which provides pressure support is fully ionized and produces no detectable absorption. The stability of such an outflow is discussed in Section~\ref{sec:discussion}.

What happens if the absorbing gas is accelerated by the radiation pressure? If a fraction $f_{\rm ac}$ of the incident $P_{\rm rad}$ leads to an acceleration of the outflow, then only a fraction of $1-f_{\rm ac}$ compresses the gas and contributes to the hydrostatic solution. The gas pressure will build up to $(1-f_{\rm ac})P_{\rm rad}$, and the ionization parameter to $U/(1-f_{\rm ac})$. However, since a fully hydrostatic solution leads to $U\simeq 0.1$ at $\tau\sim 1$, even a value as high as $f_{\rm ac}=0.9$, which implies $U\sim1$, still provides sufficient compression to produce dense enough gas which produces the observed high ionization lines. If $f_{\rm ac}=1$, then the outflowing gas will not be compressed at all. Self-consistent calculation of the outflow dynamics and ionization structure is required to derive the dynamics of the absorbing gas, as further discussed in Section~\ref{sec:discussion}.

\subsection{An analytic estimate of \Nion}\label{sec:model_approx}
RPC implies a nearly universal set of ionic columns \Nion\ in the absorber, independent of $L$ and $r$. The set of predicted \Nion\ values can be used to test the validity of the RPC mechanism. Below we provide a simple analytic estimate for \Nion.

A given ion is mostly produced in a layer which has the optimal range of $U$ values for this ion ($U_{\rm ion,2}\leq U\leq U_{\rm ion,1}$). Since this range is generally small ($\la1$~dex; e.g. \citealt{hamann_97}), the layer has an approximately constant $T$, and a given $\bar{\sigma}$. The hydrostatic solution (equation~\ref{eq:dP_dr}) for the density structure of such a layer is 
\begin{equation}
 n=n_1\exp\left(\frac{r-r_1}{l_{\rm pr}}\right), \label{eq:n_struct}
\end{equation}
where
\begin{equation}
 l_{\rm pr}=\frac{2kTc}{F_{\rm rad}\bar{\sigma}}, \label{eq:l_pr}
\end{equation}
and $r_1$ and $n_1$ are the distance and density at the illuminated face (where $U=U_{\rm ion,1}$; paper II). We assume below for simplicity that $\tau(r)\ll 1$, and that the geometrical dilution of $F_{\rm rad}$ is negligible, i.e.\ $(r-r_1)/r_1\ll1$. The ionic column for an element with abundance $f_{\rm elm}$ is
\begin{equation}
 \Nion=f_{\rm elm}\int_{r_1}^{r_2} f_{\rm ion} n\dd r,
\end{equation}
where $f_{\rm ion}$ is the ionic fraction, and $r_2$ is the distance of the layer where $U=U_{\rm ion,2}$. Changing the integration variable from $r$ to $n$, where $\dd n = \dd r\, n/l_{\rm pr}$ (equation~\ref{eq:n_struct}), yields
\begin{equation}
 \Nion=f_{\rm elm}l_{\rm pr} \int_{n_1}^{n_2}f_{\rm ion} \dd n.
\end{equation}
The BAL gas is likely to be optically thin to the ionizing radiation, as it is not observed to produce a Lyman edge (e.g.\ \citealt*{baskin_etal13}), and we can assume that $n_\gamma$ remains constant. Since $n=n_\gamma/U$, we can now integrate over $\dd U$ instead of $\dd n$, which gives
\begin{equation}
 \Nion = f_{\rm elm} l_{\rm pr} n_\gamma \int_{U_{\rm ion,2}}^{U_{\rm ion,1}}  \frac{f_{\rm ion}}{U^2} \dd U.
\end{equation}
Noting that $F_{\rm rad}/c=n_\gamma\hnu$ implies
\begin{equation}
 \Nion=\frac{2kT}{\hnu\bar{\sigma}}f_{\rm elm}\int_{U_{\rm ion,2}}^{U_{\rm ion,1}}  \frac{f_{\rm ion}}{U^2} \dd
U, \label{eq:approx_Nion}
\end{equation}
i.e.\ a universal \Nion, which depends only on the ionizing Spectral Energy Distribution (SED) and metallicity ($Z$), and is independent of $L$ and $r$. The integral can be approximated as $ \langle f_{\rm ion}\rangle /U_{\rm ion,1} $, where $ \langle f_{\rm ion}\rangle$ is the mean ion fraction in the layer, which then gives
\begin{equation}
 \Nion=\frac{2kTf_{\rm elm} \langle f_{\rm ion}\rangle }{\hnu\bar{\sigma}U_{\rm ion,1}}. \label{eq:an_Nion}
\end{equation}

Equation~\ref{eq:an_Nion} can be utilized to estimate \Nion. For example, we estimate below \Nion\ for C$^{3+}$. For $Z=Z_{\sun}$, the abundance of C is $\log f_{\rm elm}=-3.6$,\footnote{We quote the value of C abundance from \citet*{allende_prieto_etal02} which is adopted by the photoionization code \cl\ (see Section~\ref{sec:model_numerical}).} and $\langle f_{\rm ion}\rangle/U \approx0.5/0.05=10$ (e.g.\ \citealt{hamann_97}; paper II). As shown above, RPC implies $2kT/\hnu\approx 0.1$ near the H ionization front, where most of \Nion(C$^{3+}$) is produced. The value of $\bar{\sigma}$ depends on whether the absorber contains dust. For a dustless absorber, $\bar{\sigma}\approx10^{-22}$~cm$^2$, which yields $\Nion(\mbox{C$^{3+}$})\approx10^{18.4}$~\cmmt. For a dusty absorber, $\bar{\sigma}\approx10^{-21}$~cm$^2$ and $\Nion(\mbox{C$^{3+}$})\approx10^{17.4}$~\cmmt. These two values are comparable to $\Nion(\mbox{C$^{3+}$})\simeq10^{18}$ and $10^{17}$~\cmmt\ derived from photoionization calculations for a dustless and dusty absorber, respectively (Section~\ref{sec:results}).

\subsection{The numerical solutions}\label{sec:model_numerical}
We use the photoionization code \cl\ 10.00 \citep{ferland_etal98} to solve for the ionization structure of an RPC slab. We run the code with the `constant pressure' command, which requires the code to find solutions that satisfy equation~\ref{eq:dP_dr}. The contribution of trapped line emission pressure to the gas pressure is not included for technical reasons of code convergence, but it is not expected to have a significant effect (as discussed in papers II and III). We assume a H total column in the range of $19\leq\log\NHtot\leq24$. Three types of SED are adopted (see paper II), which differ in their ionizing slope \aion\ ($f_\nu\propto\nu^\alpha$) between 1~Ryd and 1~keV (912--12~\AA). The hard, intermediate and soft SED have $\aion=-1.2$, $-1.6$ and $-2.0$, respectively. We explore three values of metallicity, $Z=0.5$, 1 and 5$Z_{\sun}$, and adopt the scaling law of the metals with $Z$ from \citet{groves_etal04}. The RPC solution is independent of the assumed density at the illuminated face of the slab (papers I and II) for the specific value assumed here of $10^4$~\cmt. We use $L_{46}=1$ in our calculations to set the physical scale. The explored range in slab distance from the continuum source is 0.1--10\rdust\ for dustless models and 1--100\rdust\ for dusty models (see below). Note that for photoionized gas that absorbs in the UV, $T$ is approximately a few times $10^4$~K. This $T$ is too low for destroying dust grains by sputtering, and too high to allow grain nucleation. Thus, if the gas is dustless (or dusty) at the outflow origin, it remains such throughout the outflow.

To account for the effect of the observed BAL velocity dispersion on the radiative transport calculations, we execute the photoionization code with the `turbulence' command. The turbulence velocity $b$ parameter is set to be 2000~\kms, which is similar to the width of the typical BAL profile \citep{gibson_etal09, baskin_etal13}, and allows the escape of resonance lines. The BAL velocity dispersion is unlikely to be produced by turbulence, as the required turbulence velocity is highly supersonic (Mach number of $\sim2000/10=200$), and the gas would be shocked. Instead, the observed velocity dispersion is likely produced by an ordered velocity field. This mechanism is further discussed in Section~\ref{sec:discussion}. We make use of the `turbulence' command since this is the simplest procedure to include the effect of velocity dispersion on the escape of resonance lines in the radiative transport calculations of \cl. The turbulent pressure is not included in the total pressure of the slab, as the observed BAL velocity dispersion originates from large scale ordered motion, rather than from small scale turbulence.
 
We assume the Galactic ISM grain composition and metal depletion for dusty models, although BALQs are observed to contain dust which has an extinction law similar to that of the Small Magellanic Cloud (SMC; \citealt{sprayberry_foltz92, baskin_etal13}). The SMC grain composition is not used, since it is not one of the built-in grain types of \cl, and incorporating SMC dust into the code is beyond the scope of this study. The adopted grain composition should not affect significantly the calculation results, given the overall similarity of the Galactic and SMC extinction curves \citep{weingartner_draine01}. We scale linearly the dust to gas ratio with $Z$, and disregard sublimation of small grains at $r\ga\rdust$.

Below, we adopt models with the intermediate SED (i.e.\ $\aion=-1.6$), $Z=Z_{\sun}$ and $r=r_{\rm dust}$, unless otherwise noted.

\section{Results}\label{sec:results}

\subsection{Comparison with uniform-density models}

\begin{figure*}
 \includegraphics[width=175mm]{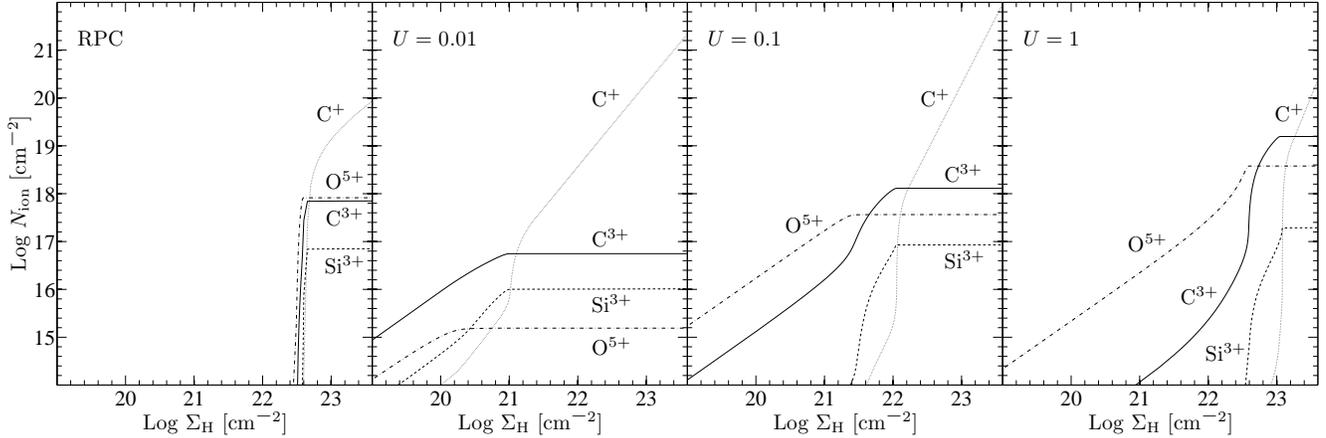}
\caption{
A comparison of \Nion\ versus \NHtot\ derived from the RPC solution for various ions (leftmost panel), and the relations derived from uniform-density solutions for different $U$ values (indicated in each of the other three panels). The RPC solution shows a sharp rise in \Nion\ to its maximal value within a small range of \NHtot\ ($\sim$0.2~dex), which occurs at $\NHtot\simeq 10^{22.5}$~\cmmt. The \Nion\ of the low-ionization ions continues to increase within the inner partially-neutral region. In contrast, the uniform-density solutions produce a gradual rise in \Nion\ over a wide range ($\sim$2 dex) of \NHtot, with a maximal value which depends on $U$.
}\label{fig:const_den_Nion_Sigma}
\end{figure*}

Figure~\ref{fig:const_den_Nion_Sigma} demonstrates the main difference between RPC models and uniform-density models. The figure compares the calculation of \Nion\ as a function of \NHtot\ for an RPC model and three uniform-density models, with $U=0.01$, 0.1 and 1. The gas is assumed to be dustless. For the RPC model, \Nion\ sharply increases by $>$3~dex in a small range of \NHtot\ ($\sim$0.2~dex), and reaches its asymptotic value for the high- and intermediate-ionization ions. The sharp increase in \Nion\ occurs at $\NHtot\simeq10^{22.5}$~\cmmt\ for ions of all ionization states. The rise in \Nion\ of low-ionization ions (see C$^+$) continues for larger \NH, but with a shallower slope. The RPC solution for \Nion\ versus \NHtot\ is unique, and is independent of the assumed $L$, $r$ and $n$ at the illuminated face of the RPC slab (papers I and II). In contrast, the uniform-density models show a gradual increase of \Nion\ with \NHtot, and the asymptotic value of \Nion\ depends on the adopted value for $U$.

\subsection{The structure of a dustless absorber}

\begin{figure}
 \includegraphics[width=84mm]{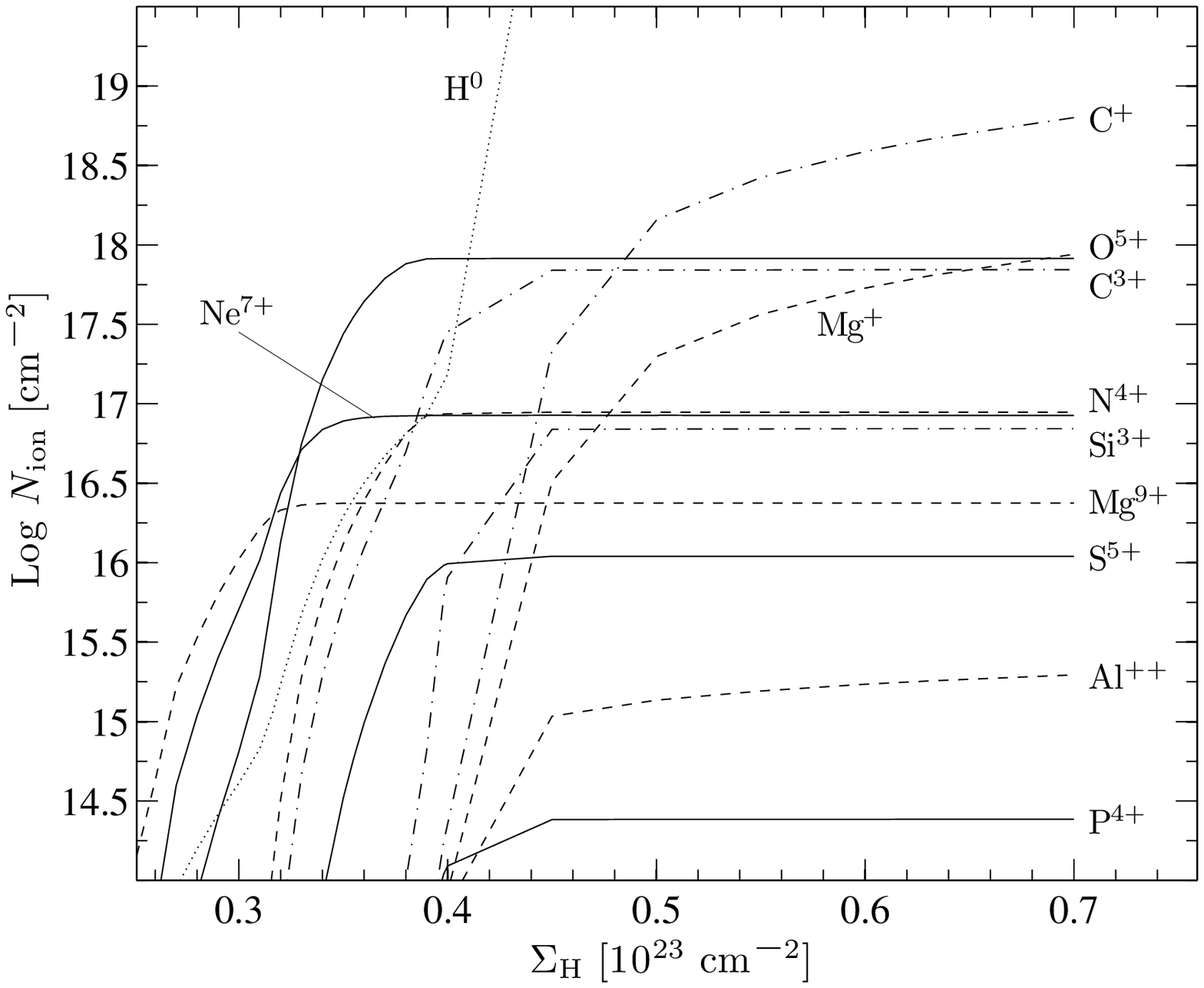}
\caption{
A zoom on \Nion\ versus \NHtot\ for the RPC model presented in Figure~\ref{fig:const_den_Nion_Sigma}, left panel. The column of most ions sharply increases at $\NHtot\simeq 0.3-0.4 \times10^{23}$~\cmmt, and reaches its asymptotic value at $\NHtot\simeq 0.5\times10^{23}$~\cmmt. The column of low-ionization ions, with a creation energy below 1~Ryd (i.e.\ C$^+$ and Mg$^+$), continues to increase above this value of \NHtot, as they are present in the partially-neutral constant-density region beyond the H ionization front in the slab.
}\label{fig:NOdust_Nion_Sigma}
\end{figure}

Figure~\ref{fig:NOdust_Nion_Sigma} presents a more detailed view of \Nion\ as a function of \NHtot\ for a variety of ions. The calculated \Nion\ of all ions sharply increases from $<10^{14}$~\cmmt\ at $\NHtot<0.2\times10^{23}$~\cmmt\ to the asymptotic value at $\NHtot\simeq0.3 - 0.4\times10^{23}$~\cmmt. The steep gradient of \Nion\ can be explained as follows. For a given ion, \Nion\ accumulates in the range of \NHtot\ which corresponds to the range of $U$ that is optimal for the creation of this ion. In the range of \NHtot\ where the slab is optically thin to the ionizing radiation, $n_{\gamma}$ is nearly constant, and thus $U$ is set by $n$. For an RPC gas, $n$ rises exponentially with depth (equation~\ref{eq:n_struct}), which produces the steep increase of \Nion\ with \NHtot\ towards the optimal $U$ region. Slightly deeper, $U$ drops steeply below the optimal values, and \Nion\ reaches the asymptotic value.

The asymptotic value of $\log \Nion$ is $\simeq$18 for O$^{5+}$ and C$^{3+}$, and 17 for N$^{4+}$ and Si$^{3+}$. For low-ionization ions, with creation energy $<$1~Ryd (i.e.\ C$^+$ and Mg$^+$), \Nion\ does not reach an asymptotic value, and continues to increase with a shallow slope in the partially ionized region at $\NHtot > 0.5\times10^{23}$~\cmmt, where the density remains uniform. At $\NHtot = 0.5\times10^{23}$~\cmmt, the calculated $\log\Nion$ of C$^+$ and Mg$^+$ is $\simeq$18 and 17.3, respectively, which produces strong Low ionization Broad Absorption Lines (LoBALs). Since the fraction of LoBALQs in the BALQ population is only a few per cent \citep{trump_etal06, allen_etal11}, there should be a mechanism that caps the absorber at $\NHtot <0.4\times10^{23}$~\cmmt\ so that \Nion\ of the low-ionization ions does not build up to significant values ($>10^{16}$~\cmmt; see discussion in Section~\ref{sec:discussion}).

\begin{figure}
 \includegraphics[width=69.4mm]{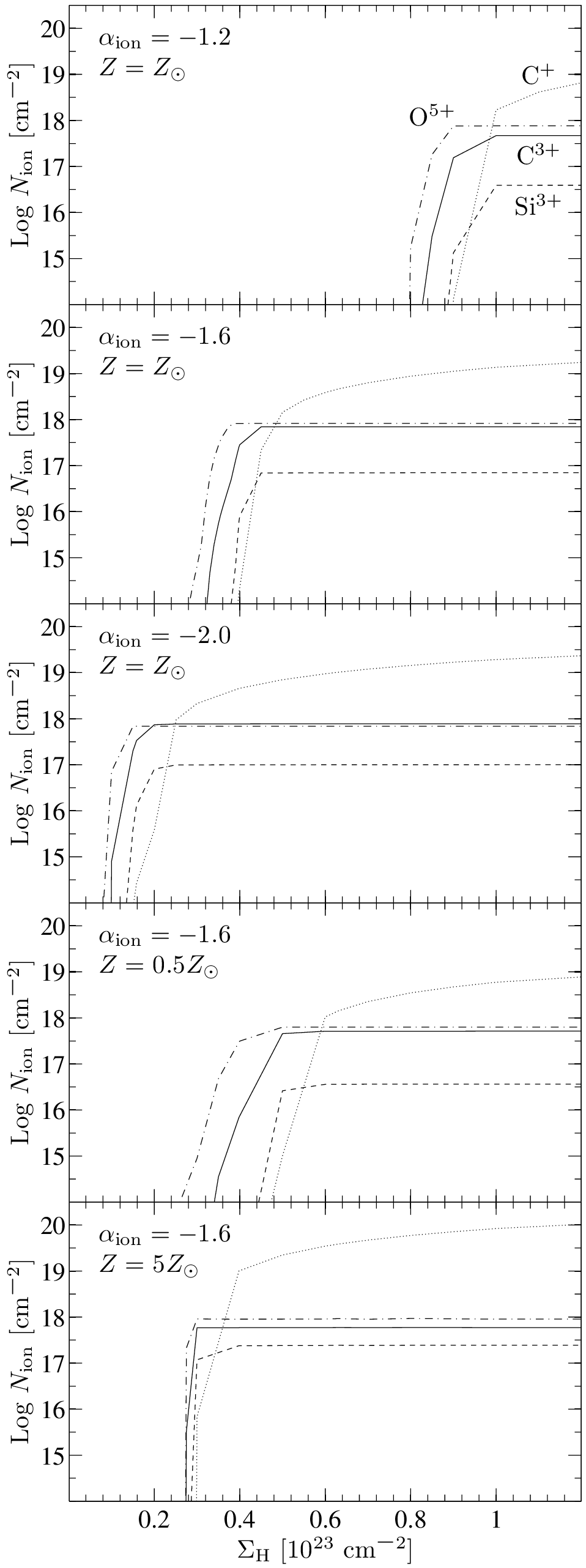}
\caption{
The effect of \aion\ and $Z$ on the solution for \Nion\ versus \NHtot\ in a dustless RPC slab. The values of \aion\ and $Z$ are noted at each panel. The value of \NHtot\ where \Nion\ sharply increases, depends mostly on \aion, increasing from $\simeq0.1\times10^{23}$~\cmmt\ for $\aion=-2.0$ to $\simeq0.8\times10^{23}$~\cmmt\ for $\aion=-1.2$. The width of the transition range is proportional to $Z^{-1}$.
}\label{fig:NOdust_all_Nion_Sigma}
\end{figure}

Figure~\ref{fig:NOdust_all_Nion_Sigma} presents the dependence of \Nion(\NHtot) on \aion\ and $Z$ for various ions. The upper three panels present \Nion(\NHtot) for $\aion=-1.2$, $-1.6$ and $-2.0$, for $Z=Z_{\sun}$. The sharp rise in \Nion\ occurs at $\NHtot\simeq0.9$, 0.3 and $0.1\times10^{23}$~\cmmt\ for $\aion=-1.2$, $-1.6$ and $-2.0$, respectively. This trend occurs because a softer ionizing SED (i.e.\ a more negative \aion) implies a smaller $n_{\gamma}$ at a given $L_{\rm bol}$, and therefore requires a lower $n_e$, or a smaller \NH, to reach the optimal $U$. The width of the transition range, $\simeq 0.1\times 10^{23}$~\cmmt, is roughly independent of \aion. The two bottom panels present models with $Z=0.5$ and 5$Z_{\sun}$, for $\aion=-1.6$. The transition range centre is roughly independent of $Z$, and is located at $\NHtot\simeq 0.3\times10^{23}$~\cmmt. The width of the range decreases from 0.2, through 0.1, to $\simeq0.02\times10^{23}$~\cmmt\ for $Z=0.5$, 1 and 5$Z_{\sun}$, respectively. The width is therefore proportional to $Z^{-1}$, as expected since the metals dominate the opacity in this range of $U$, and a given $\tau$ is required to reach the optimal $U$.

To summarize, Figure~\ref{fig:NOdust_all_Nion_Sigma} indicates that \aion\ mainly controls the centre of the transition range of \NHtot, while $Z$ mainly controls its width. 

\begin{figure}
 \includegraphics[width=84mm]{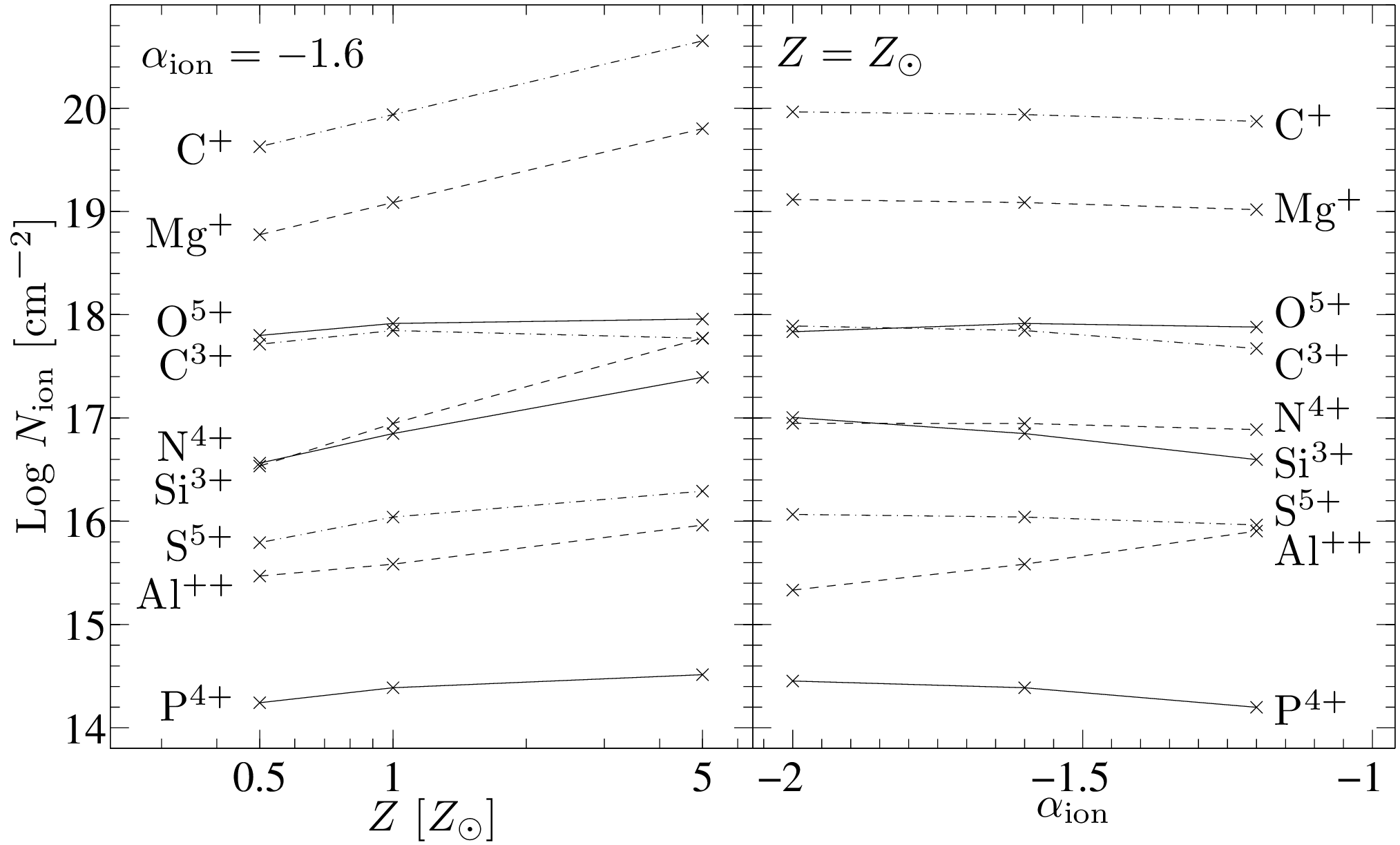}
\caption{
The asymptotic value of \Nion\ of various ions versus $Z$ (left panel) and \aion\ (right panel). The slab is dustless and has $\NHtot=4\times 10^{23}$~\cmmt. Left panel presents models with $0.5\le Z/Z_{\sun}\le 5$ and $\aion=-1.6$; right panel presents $-2\le \aion\le -1.2$ and $Z/Z_{\sun}=1$ (each cross marker denotes a specific photoionization model run). The value of $Z$ has a prominent ($\ga$1~dex) effect on the low-ionization ions, and on N$^{4+}$. The adopted \aion\ has a significant effect only on Al$^{++}$ ($\simeq$0.7~dex), and a negligible effect on most other ions.
}\label{fig:NOdust_Nion_Z_and_alpha}
\end{figure}

Figure~\ref{fig:NOdust_Nion_Z_and_alpha} presents the dependence of the asymptotic value of \Nion\ on $Z$ and \aion. We use $\NHtot=4\times10^{23}$~\cmmt, which is well above \NHtot\ for which \Nion\ of all ions (except C$^+$ and Mg$^+$) reaches the asymptotic value (Fig.~\ref{fig:NOdust_all_Nion_Sigma}). The left panel presents the $Z$ dependence for $Z=0.5-5$,  for $\aion=-1.6$. As expected, \Nion\ increases with $Z$. The increase is linear for the low-ionization ions (C$^+$ and Mg$^+$), but is smaller than linear for the higher-ionization ions. The sub-linear increase results from the decreasing width of the transition \NHtot\ with increasing $Z$ (Fig.~\ref{fig:NOdust_all_Nion_Sigma}), which compensates part of the increase of element abundance with $Z$. The only exception is the N$^{4+}$ column where the dependence is steeper than linear, as expected since the N abundance $\propto Z^2$ (Section~\ref{sec:model_numerical}). Here as well, the rise in \Nion\ of N$^{4+}$ is lower than $Z^2$ due to the decrease in the width of the transition \NHtot.

The right panel of Fig.~\ref{fig:NOdust_Nion_Z_and_alpha} presents the dependence of \Nion\ on \aion\ for values in the range $-2.0$ to $-1.2$, for $Z=Z_{\sun}$. The value of \Nion\ is nearly independent of \aion\ for most ions, and typically varies by less than 0.5~dex for the ions which show either positive or negative dependence.

\begin{figure}
 \includegraphics[width=84mm]{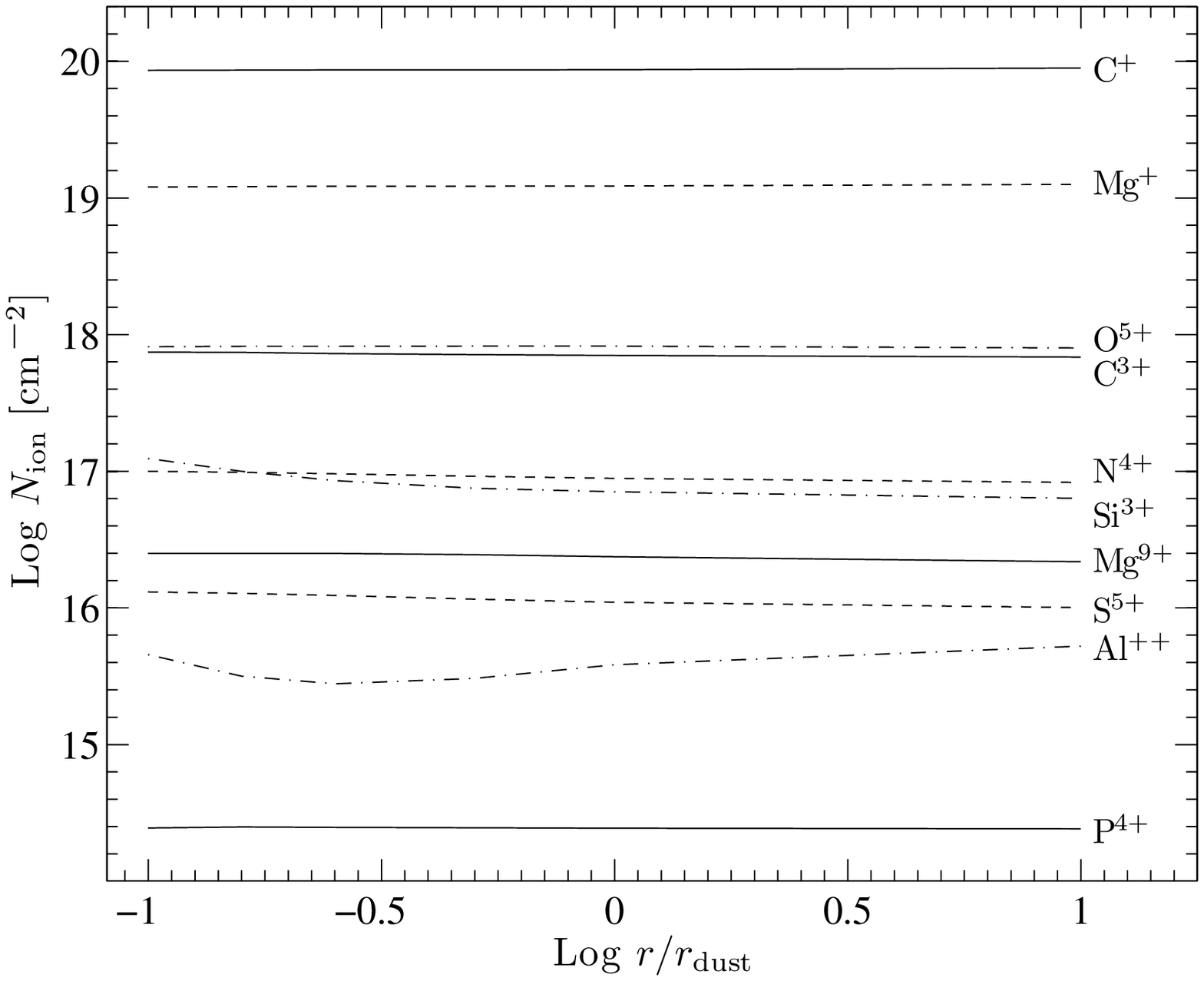}
\caption{
The dependence of the asymptotic value of \Nion\ on the distance from the ionizing source, measured in units of $r/r_{\rm dust}$. The slab is dustless with $\NHtot=4\times 10^{23}$~\cmmt. The ionic column is independent of $r$, over the explored range of 2~dex in $r$. This demonstrates the universality of the RPC solution, which is not affected by the distance from the AGN.
}\label{fig:NOdust_Nion_R}
\end{figure}

Figure~\ref{fig:NOdust_Nion_R} presents the asymptotic \Nion\ of various ions as a function of distance. It shows that \Nion\ is nearly independent of distance, despite the factor of 100 in distance, and therefore the factor of $10^4$ in $n_e$ at a given $\tau$. This reflects the nearly universal $U(\tau)$ RPC solution, which is almost independent of distance (e.g.\ paper I). Thus, RPC provides a clear signature in the distribution of \Nion\ values, which depends significantly only on the gas metallicity. The distribution is nearly independent of the gas environment, distance from the ionizing source, and the illuminating SED. However, one should keep in mind the significant uncertainty concerning the validity of the hydrostatic solution, as further discussed is Section~\ref{sec:discussion}.

\subsubsection{The predicted UV versus X-ray absorption spectrum}
\begin{figure*}
 \includegraphics[width=84mm]{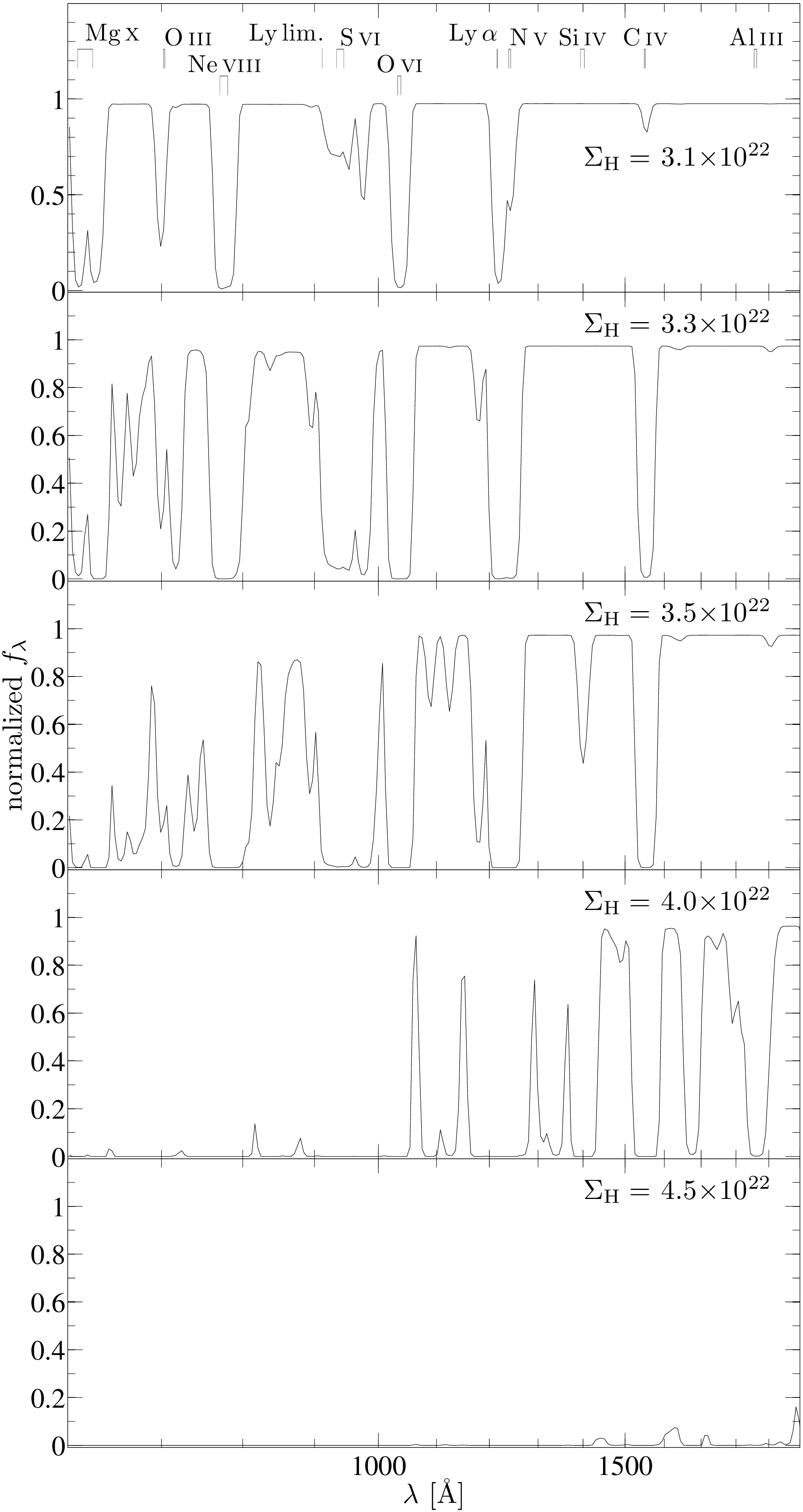}
 \includegraphics[width=84mm]{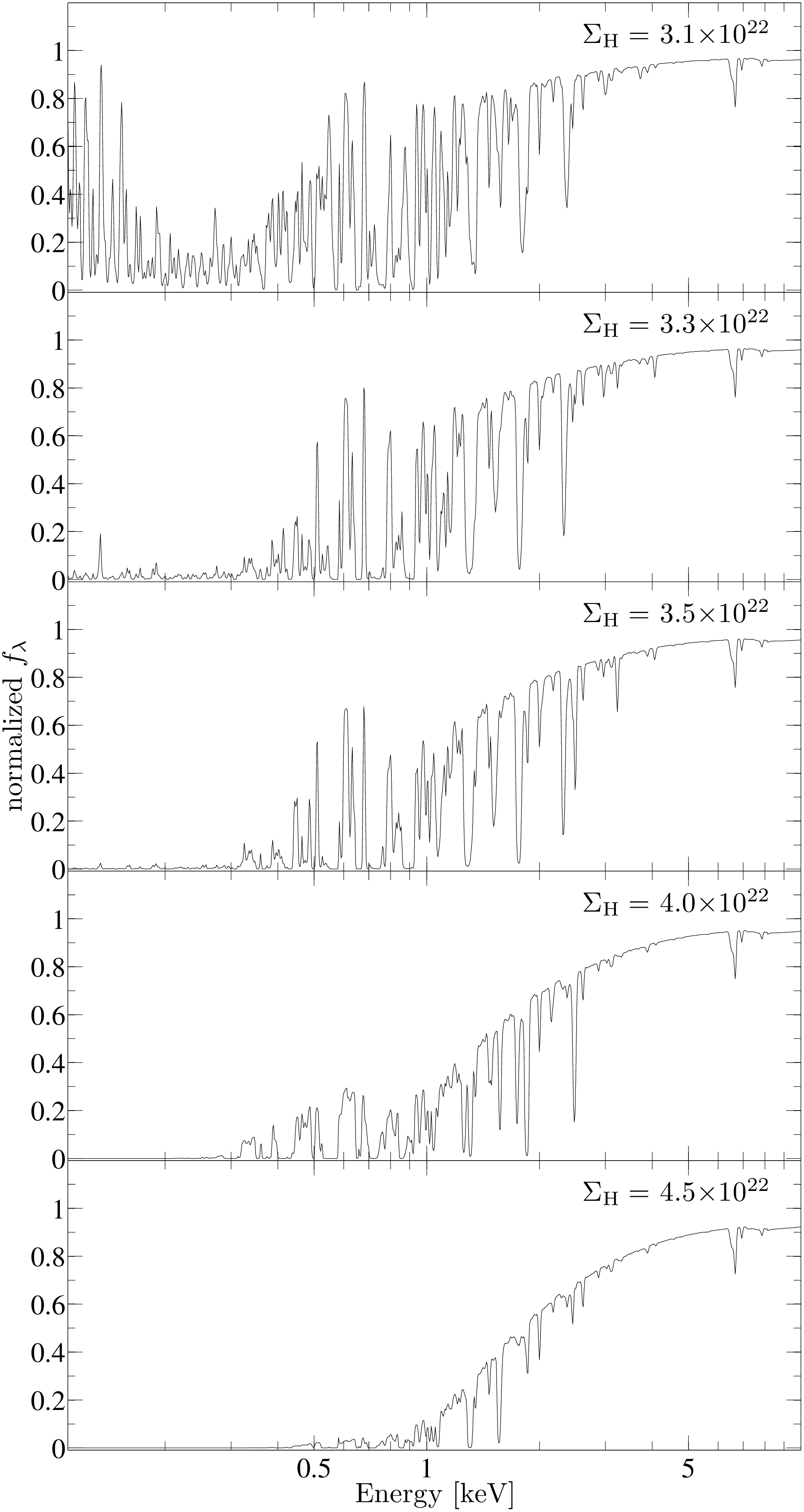}
\caption{
The absorption spectrum of a dustless RPC slab in the UV (left panels) and in the X-ray range (right panels), as calculated by \cl. Five different values of \NHtot\ are adopted, as indicated in each panel. The three lower values correspond to different strengths of \CIV\ absorption (left panels), from marginal to heavily saturated. The two highest \NHtot\ values produce LoBALs and a nearly fully absorbed UV spectrum. The spectral sampling is in increments of 1500~\kms, and the assumed absorber FWHM is 3300~\kms. The slab absorption at 2~keV (right panels) is relatively small ($\le0.25$~dex) for $\NHtot\le4\times10^{22}$~\cmmt, indicating that the steep \aox\ observed in HiBALQs is not caused by the UV absorbing gas.
}\label{fig:synth_flux}
\end{figure*}

Figure~\ref{fig:synth_flux} compares the predicted absorption spectrum in the UV and in the X-ray regime for five values of \NHtot, as derived from the \cl\ calculations. The values of \NHtot\ of 3.1, 3.3, 3.5, 4 and $4.5\times 10^{22}$~\cmmt, are chosen such that a marginal, a saturated and a heavily saturated \CIV\ absorption is produced for the first three values. The last two values produce LoBALs and a nearly fully absorbed UV.  The flux at $E=2$~keV, which is used to measure \aox, is not significantly affected by the absorption ($\leq0.25$~dex) for $\NHtot\leq 4\times10^{22}$~\cmmt. A similar result was found by \citet{hamann_etal13} for a uniform-density absorber. This implies that the measured \aox\ for High ionization BALQs (HiBALQs) is a good tracer of the intrinsic SED. The column required to produce significant X-ray absorption above 2~keV extinguishes the UV shortward of 2000~\AA\ completely. A dusty absorber produces more significant UV absorption at a given $\NHtot$ (see below), and therefore implies a weaker X-ray absorption. We further discuss the implications of the X-ray absorption spectrum in Section~\ref{sec:discussion}.

\subsection{Dusty absorber}

\begin{figure*}
 \includegraphics[width=150mm]{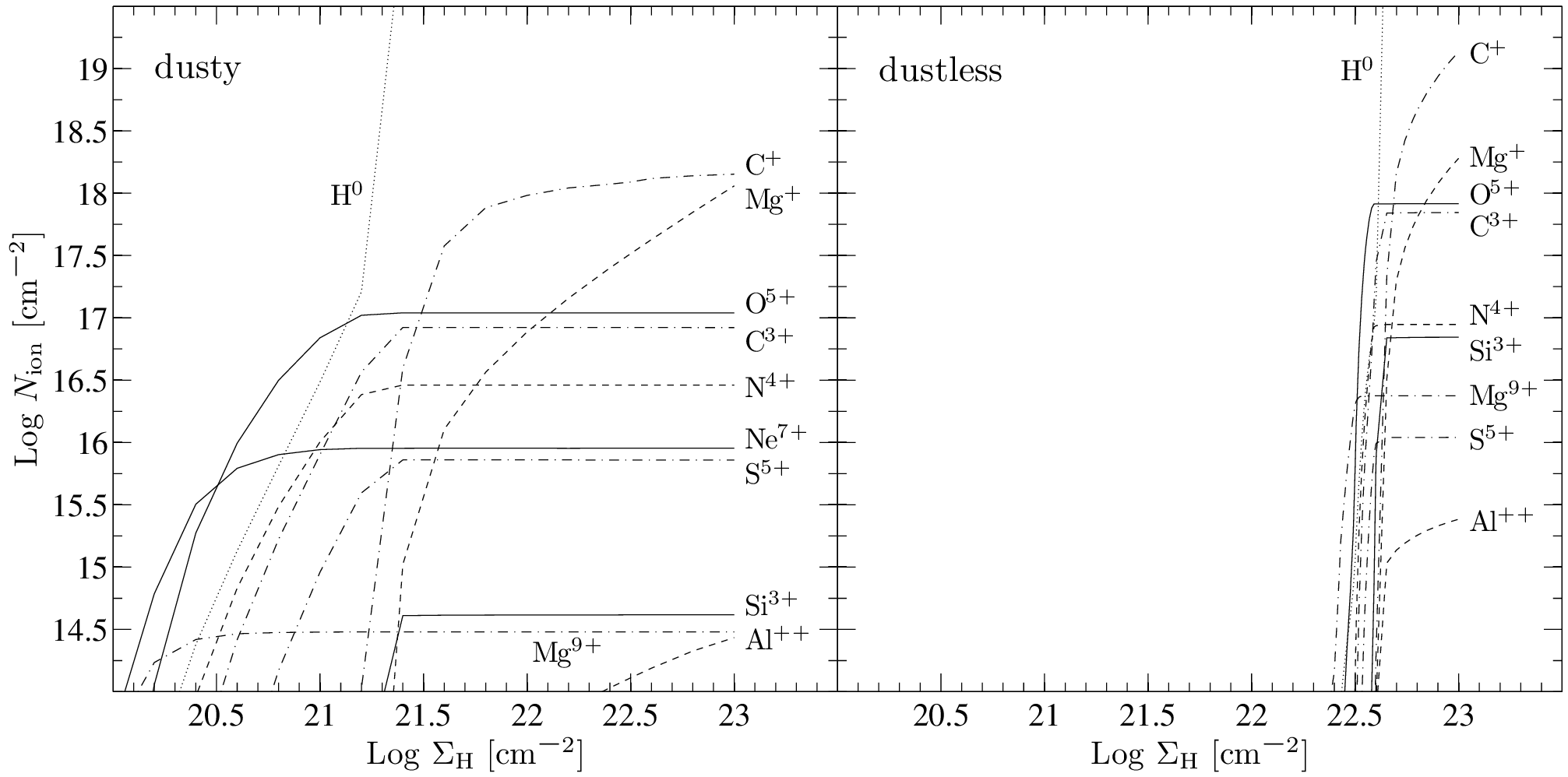}
\caption{
A comparison between a dusty and dustless RPC slab. Note that \NHtot\ is on a logarithmic scale (cf.\ Fig.~\ref{fig:NOdust_Nion_Sigma}, where the scale is linear). The increase of \Nion\ with \NHtot\ is significantly more gradual for the dusty model (left panel) compared to the dustless model (right panel). For the dusty model, high- and intermediate-ionization ions reach their asymptotic \Nion\ at $\NHtot\simeq 10^{21.5}$~\cmmt, which is smaller by 1~dex compared to the dustless model. The asymptotic value of \Nion\ in the dusty model is smaller by $\ga$1~dex than in the dustless model for most ions. In particular, the asymptotic value of \mbox{\Nion(Si$^{3+}$)} is $<10^{15}$~\cmmt\ in the dusty model, compared to $10^{16.8}$~\cmmt\ in the dustless model, since Si is heavily depleted into grains in the Galactic ISM composition which we adopt for the dusty model. 
}\label{fig:dust_Nion_Sigma}
\end{figure*}

Figure~\ref{fig:dust_Nion_Sigma} compares the relation between \Nion\ and \NHtot\ for a dusty and a dustless slab. There are two main differences. First, in dusty gas the asymptotic value of \Nion\ is reached at $\NHtot\simeq 10^{20.5}-10^{21.5}$~\cmmt\ depending on the ion ionization state, compared to $\NHtot\simeq10^{22.5}$~\cmmt\ for dustless gas. Second, the rise in \Nion\ to its asymptotic value occurs over a range of $\Delta\NHtot\approx1$~dex in dusty gas, compared to a much sharper rise which occurs over $\Delta\NHtot\approx0.1-0.2$~dex in dustless gas. The first effect is expected since $\bar{\sigma}$ is larger by $\sim$1~dex in dusty gas than in dustless gas ($\approx10^{-21}$ for dust versus $\approx10^{-22}$~cm$^2$ or lower for the gas). The more gradual increase of \Nion\ in dusty gas results from the constant dust opacity of dusty gas, in contrast with the ionization dependent gas opacity which increases with increasing gas density in dustless gas, and leads to a steep rise of $n$ with \NHtot\ (paper I).

Dust is the dominant absorber of the ionizing radiation for $U>10^{-2}$ \citep{netzer_laor93}. In RPC gas, $U\sim 0.1$, and therefore most ($\sim 80$ per cent) of the ionizing photons are absorbed by dust and converted to thermal dust emission (paper I), rather than ionize the gas. This leads to the order of magnitude reduction in \Nion\ for most ions (for the very high-ionization Mg$^{9+}$ ion the column is reduced by $\simeq$2~dex). In some cases the reduction is smaller, due to the modified ionization structure (S$^{5+}$, Mg$^+$ and N$^{4+}$ decrease by $\simeq$0.2--0.5~dex). Some of the elements are expected to be heavily depleted into grains, in particular Si, which leads to a reduction in the asymptotic \Nion\ of Si$^{3+}$ by 2~dex.

\begin{figure}
\includegraphics[width=69.4mm]{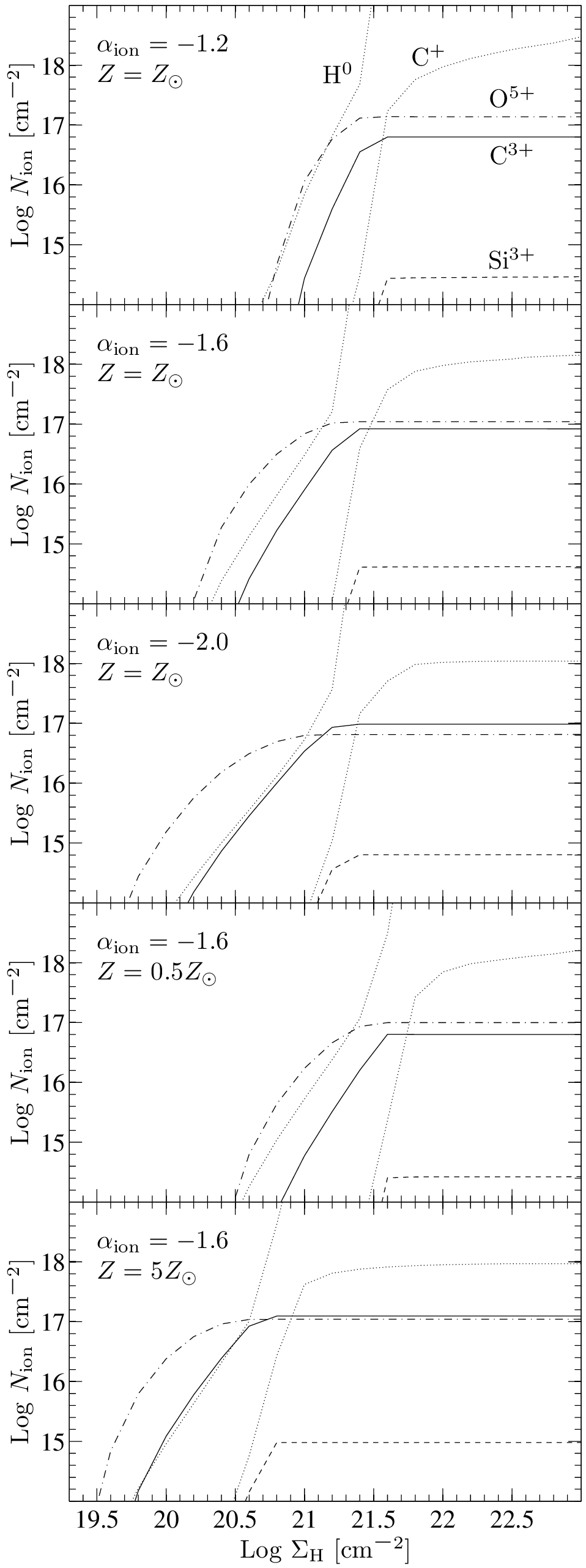}
\caption{
The same as Fig.~\ref{fig:NOdust_all_Nion_Sigma}, for a dusty RPC slab. Note that \NHtot\ is on a logarithmic scale (cf.\ Fig.~\ref{fig:NOdust_all_Nion_Sigma}, where the scale is linear). Varying the assumed \aion\ and $Z$ has a similar effect on \NHtot\ at which \Nion\ starts to increase. A steeper \aion\ (higher $Z$) shifts this \NHtot\ to a lower value compared to a shallower \aion\ (lower $Z$). The width of the transition range here also scales as $Z^{-1}$.
}\label{fig:dust_all_Nion_Sigma}
\end{figure}

Figure~\ref{fig:dust_all_Nion_Sigma} presents the dependence of the transition \NHtot\ and the asymptotic \Nion\ on \aion\ and $Z$ in a dusty absorber. The transition \NHtot\ is smaller for a softer \aion, as in the case of a dustless slab (Fig.~\ref{fig:NOdust_all_Nion_Sigma}). This results from the increasing dust $\bar{\sigma}$ for the softer SED, which is more strongly dominated by the UV emission. A rise in $Z$ lowers the transition \NHtot, as expected given the increase of dust content with $Z$, which implies a rise in $\bar{\sigma}$. The asymptotic \Nion\ are effectively independent of \aion, and show some rise with $Z$, which is different for different ions. The expected reddening associated with \Nion\ is discussed in Section~\ref{sec:discussion}.

\section{Discussion}\label{sec:discussion}

As shown above, RPC may solve the overionization problem of BAL outflows, by naturally explaining the small filling factor of the absorbing gas. The small filling factor solution is necessary given the observed absence of a large enough suppression of the ionizing continuum along our line of sight \citep{hamann_etal13}, as required by the X-ray shield mechanism to prevent overionization \citep{murray_etal95, chelouche_netzer01}. The compression of the gas, which leads to the small filling factor, is an inevitable effect in photoionized gas, and leads to the observed ionization state of the gas. There are no free parameters involved in the solution. The outflowing gas forms thin sheets or filaments transverse to the outflow, as diffuse gas which potentially rises from the disc (likely a disc wind) is compressed along the radial direction, while maintaining its original extent in the transverse direction.

Furthermore, RPC leads to a wide distribution of ionization states in the outflowing gas, due to the density gradient in the radial direction. This explains the observed similar absorption profiles from a wide range of ionization levels, which lead to the suggestion of a multiphase outflow \citep{everett_etal02}.

A robust prediction of RPC is the value of \Nion\ for various ions. The predicted set of \Nion\ values is effectively independent of distance, is weakly dependent on \aion, and shows some dependence on $Z$. The predicted values can be used to test the validity of RPC. The  predicted column of the UV absorbing layer is $\simeq 10^{22}$~cm$^{-2}$ (Fig.~\ref{fig:NOdust_Nion_Sigma}). Generally, the required column to produce significant absorption over a range of say 3000~\kms\ by a specific ion is $\simeq 3\times 10^{16}$~cm$^{-2}$, for the typical oscillator strength of $\sim 0.5$ for resonance lines. Thus, the most abundant elements, C, N, O and Ne (X/H$\sim 10^{-4}$), are expected to produce highly saturated broad absorption lines. The somewhat less abundant elements of Mg, Si, S and Fe (X/H$\sim 10^{-5}$) can also produce saturated absorption. The lower abundance elements of Na, Al, Ar and Ca (X/H$\sim 3\times 10^{-6}$) are not expected to produce strong broad absorption lines; and elements like P, Cl and K can produce detectable absorption only if narrow.

The above estimate is valid for HiBALQs. In LoBALQs the H ionization front is passed, the total absorber column is a few times larger, and the column of low-ionization ions (e.g.\ Mg$^+$ and C$^+$) is correspondingly higher (Figure~\ref{fig:NOdust_Nion_Sigma}). Furthermore, the column of the metals is linear with $Z$ (Figure~\ref{fig:NOdust_Nion_Z_and_alpha}), and thus high metallicity LoBALQs can show absorption of low-ionization lines of rather rare elements (X/H$\sim 10^{-7}$). Comparison with observations is described below (Section~\ref{sec:compare_obs}).

\subsection{The highly ionized layer}
As shown in Fig.~\ref{fig:NOdust_Nion_Sigma}, the UV absorbing layer is reached only after a column of $\simeq 3\times 10^{22}$~cm$^{-2}$ of highly ionized gas is passed. This corresponds to an electron scattering optical depth of $\tau_{\rm es}=0.02$, which yields $P_{\rm gas}=0.02P_{\rm rad}$. At this optical depth $T\simeq 10^5$~K (fig.~2 in paper I), and therefore $U\simeq 50$ (eq.~\ref{eq:U_vs_tau_T}), which is low enough to allow the gas to absorb in the UV. This `compressing layer' is required not to shield the gas from the ionizing radiation, but to allow the gas to become dense enough to start absorbing in the UV.

The width of the compressing layer $\Delta r$ cannot be larger than $r$. What are the conditions required to allow $\Delta r/r<1$? The density in this layer grows exponentially (eq.~\ref{eq:n_struct}) over a length scale $l_{\rm pr}$ (eq.~\ref{eq:l_pr}). The column density of a layer of thickness $\Delta r$, starting with a surface density $n_{\rm s}$, is 
\begin{equation}
 \NH=l_{\rm pr}n_{\rm s} \exp\left(\frac{\Delta r}{l_{\rm pr}}\right) .
\end{equation}
The value of $\Delta r$ is therefore set by $l_{\rm pr}$ and $n_{\rm s}$. Assuming $T\simeq 10^6$~K (e.g.\ fig.~2 in paper I) and $\sigma=\sigma_{\rm es}$,  and using 
\begin{equation}
F_{\rm rad}=8.4\times 10^7L_{\rm 46}r_{\rm pc}^{-2} \mbox{~\ergs~\cmmt},
\end{equation}
yields 
\begin{equation}
l_{\rm pr}=0.05L_{\rm 46}^{-1}r_{\rm pc}^{2} \mbox{~pc} . 
\end{equation}
There is therefore a minimal value for $n_{\rm s}$ which allows the slab to reach $\NH\simeq 3\times 10^{22}$~\cmmt\ within $\Delta r$. The minimal value is 
\begin{equation}
 n_{\rm s}=2\times 10^5 L_{\rm 46}r_{\rm pc}^{-2} \exp\left(\frac{-\Delta r}{l_{\rm pr}}\right) \mbox{~\cmt}.
\end{equation}
Thus, if $\Delta r/l_{\rm pr}> 10$, the required column can be reached even for $n_{\rm s}$ as low as $1$~\cmt, and the build up of the compressing layer by RPC is unavoidable. However, for $\Delta r/l_{\rm pr}< 1$, compression is not significant. This occurs for 
\begin{equation}
r_{\rm pc}>20L_{\rm 46} ,
\end{equation}
or 
\begin{equation}
\frac{r_{\rm pc}}{r_{\rm dust}}>100L_{\rm 46}^{1/2} .
\end{equation}
If the absorber is dusty, then $\bar{\sigma}\sim 10^3\sigma_{\rm es}$, the value of $l_{\rm pr}$ is correspondingly smaller and compression will be effective out to $20L_{\rm 46}$~kpc. 

Given the small $\tau$ and the high ionization of this compressing layer, dusty or dustless, it will be largely unobservable. The transition region in this layer, from the fully ionized region at $U>1000$, to the UV absorbing region at $U<10$, is a layer of highly but not fully ionized gas. This layer produces absorption edges and lines in the soft X-ray regime, e.g.\ by O$^{7+}$ and O$^{8+}$ ions, commonly observed in Seyfert galaxies. The observational signature of RPC in the soft X-ray regime is discussed in paper III. Here we just mention in passing two implications which are relevant for X-ray observations. First, the RPC absorber produces negligible absorption at 2~keV for $\NHtot\leq 4 \times 10^{22}$~\cmmt\ (Fig.~\ref{fig:synth_flux}), and is likely not responsible for the steeper values of \aox\ reported for HiBALQs compared to non-BALQs \citep{gallagher_etal06}. The contribution of the RPC absorber to $\Delta\aox$ is $\simeq-0.1$ for HiBALQs (Fig.~\ref{fig:synth_flux}), compared to a typical value of $\Delta\aox\simeq-0.5$ \citep{gallagher_etal06}. This implies that HiBALQs typically have an \emph{intrinsically} steeper \aox, as indeed indicated by the weaker \HeII\ $\lambda$1640 emission in HiBALQs, which is sensitive to the shape of the ionizing SED \citep{baskin_etal13}. Second, HiBALQs should present a complex (line) absorption in the X-ray, with \NHtot\ of a few times $10^{22}$~\cmmt, as indeed observed (e.g.\ \citealt{gallagher_etal02, piconcelli_etal10}).

We note that the highly ionized surface layer may not be present, if the absorbing gas originates in dense gas clouds which enter the radiation field (say from a disc wind). If $n_{\rm s}$ is large enough to produce say $U<10$ already at the surface, then the RPC solution remains the same, but just lacks the higher $U$ surface layers (e.g.\ paper II, fig.~2 there).

\subsection{The UV absorbing layer}
How thick is the UV absorbing layer? Only a small fraction of $\sim 10$ per cent of BALQs are LoBALQs which show absorption by ions of low ionization, such as \MgII\ and \AlIII\ \citep{reichard_etal03, trump_etal06, gibson_etal09, allen_etal11}. The RPC solution shows a sharp transition to a low ionization region at $\NH>4\times 10^{22}$~\cmmt, where the H$^0$ column sharply rises. Observations indicate that in regular high-ionization BALQs the outflow is optically thin at the Lyman limit \citep{baskin_etal13}, which implies $\NH<3.8\times 10^{22}$~\cmmt. However, for a slightly smaller $\NH<3\times 10^{22}$~\cmmt, no significant UV absorption lines are produced. What produces such a fine tuning of \NH\ in a fair fraction of AGN which are HiBALQs (3--30 per cent, depending on the \HeII\ emission strength and reddening; \citealt{baskin_etal13})?  

\begin{figure*}
 \includegraphics[width=84mm]{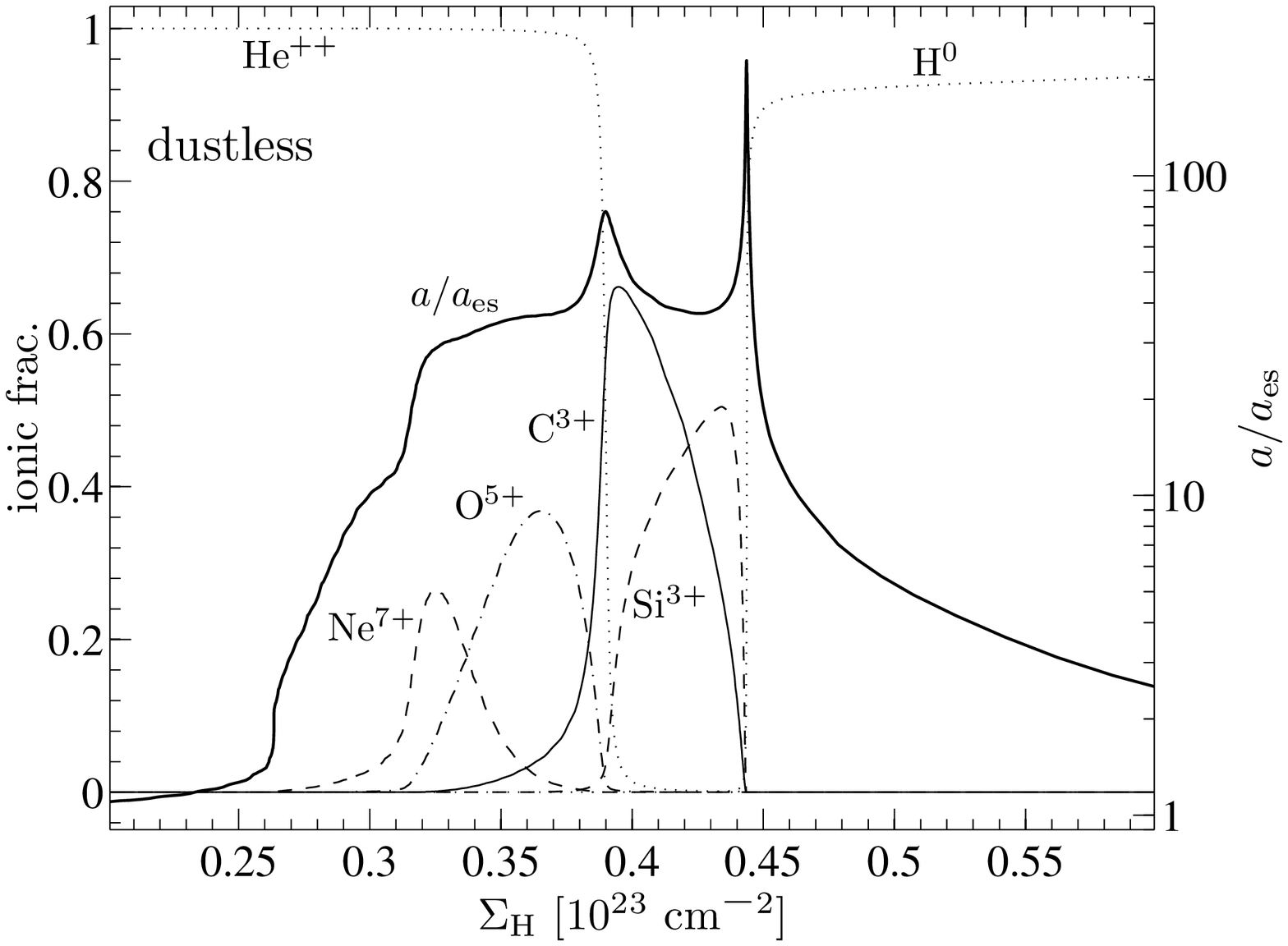}
 \includegraphics[width=84mm]{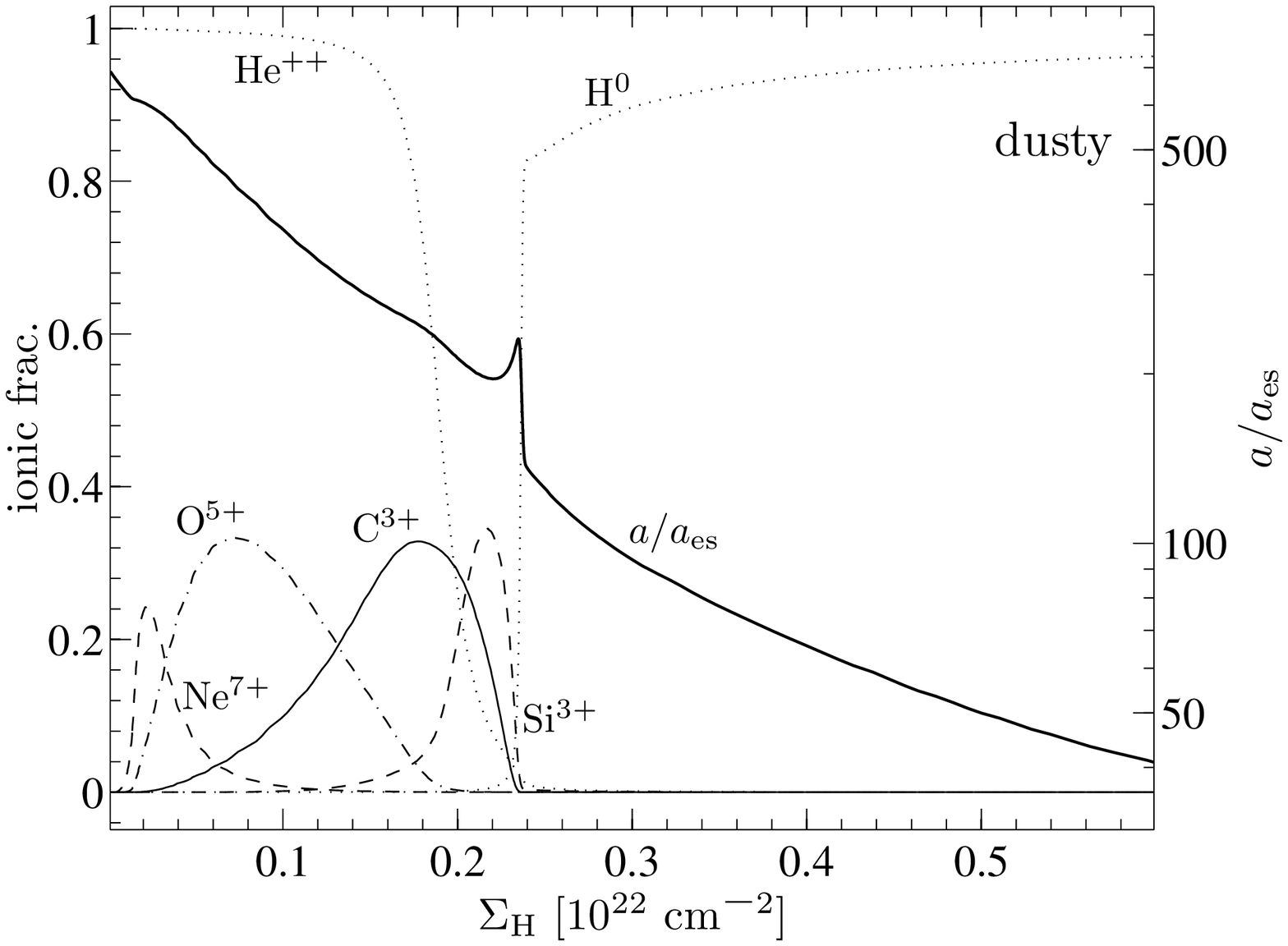}
\caption{
The calculated ionic fraction and local normalized acceleration $a/a_{\rm es}$ versus \NH\ for a dustless (left panel) and dusty RPC slab (right panel). The local acceleration is normalized by $a_{\rm es}=\sigma_{\rm es}L/4\pi r^2 c m_{\rm p}$. For the dustless slab, the two peaks of the acceleration curve correspond to the He$^+$ and the H ionization fronts. For the dusty slab, $a$ monotonically decreases with \NH, since dust is the main source of opacity, and $F_{\rm rad}$ that reaches a given layer decreases monotonically with \NH.
}\label{fig:slab_acc}
\end{figure*}

An answer to the above question may come from hydrodynamical + photoionization modelling of the outflows. To gain some insight, we present in Figure~\ref{fig:slab_acc} the force per unit mass, i.e.\ the acceleration $a$, exerted by the absorbed radiation as a function of depth into the cloud. The left panel shows the dustless gas model. Close to the surface the gas is fully ionized and $a=a_{\rm es}$, the acceleration expected from pure electron scattering. Deeper into the gas the opacity builds up, and thus $a$ increases. A peak is present near the He$^+$ ionization front, with another sharper peak near the H ionization front. In the hydrostatic case, $a$ is balanced by the gas pressure gradient. The outflowing gas is likely fed by a vertical disc wind (e.g.\ \citealt{czerny_hryniewicz11}). Once the wind is high enough above the disc, it becomes exposed to the central ionizing radiation, which produces a radial outflow. Since $a$ increases inwards as $r^{-2}$, a faster outflow is generated at smaller $r$. The inner outflow compresses the gas lying further outside which is moving at a lower velocity. The outflow is therefore compressed in the radial direction. Before the hydrostatic solution is reached the gas is radially accelerated. The plot in Fig.~\ref{fig:slab_acc} may be proportional to $a$ when the gas is on its way to reach the hydrostatic solution. The peaks in $a$ may lead to a sharp change in the local wind structure, and potentially a dynamical instability (e.g.\ Rayleigh-Taylor instability when dense gas is accelerated to a lower density gas). This may set the maximal outflowing column, as observed in HiBALQs. The exact time-dependent dynamics of the compression by radiation pressure, and possibly related instabilities, require detailed numerical simulations.

The right panel of Fig.~\ref{fig:slab_acc} shows $a$ for dusty gas. Close to the surface the opacity is dominated by dust, which gives a factor of $\sim 10^3$ enhancement in $a/a_{\rm es}$. Once the dusty gas becomes optically thick, the radiation becomes extinct and the gas falls back into the disc, e.g.\ the failed wind scenario suggested by \citet{czerny_hryniewicz11}. Only a small feature is seen near the H front, so the resulting structure may be less susceptible to an instability which may limit the absorbing column.

We note that LoBALQs always show strong absorption of high-ionization lines. There are no LoBALQs where a HiBAL component can be excluded. This agrees with RPC, as a high-ionization layer must always be present in front of the lower-ionization layer. This is in contrast with uniform-density models for LoBALQs, where a dense absorber with only low-ionization lines is a valid possibility.

\subsection{Possible contribution to the BLR emission}
Is significant line emission expected from the BAL outflow? Scattering of resonance lines is inevitable, and was used by \citet*{hamann_etal93} to exclude CF=1 in BALQs. However, in addition to resonant scattering the outflowing gas is expected to cool by line emission. The emission per unit area scales as $\propto \NH n$. In both the BLR gas and in the absorbing gas one gets $\NH\sim 10^{22}$~\cmmt\ for the UV emitting/absorbing layer. But, if the outflow is diffuse, $n$ is lower by a factor of $\sim 10^4-10^5$ than the BLR value, and thus the relative emission from the outflow is negligible. RPC implies that $n$ is comparable in both regions (paper II), and thus the emission from the outflowing gas may not be negligible. In HiBALQs, the wind is optically thin at the Lyman limit, so the total emission per unit area is smaller than expected from the BLR. However, the difference may be small for the higher-ionization lines, formed well before the H ionization front. In LoBALQs, a H ionization front is clearly present, so the emission per unit area is the same as for the BLR gas. In fact, since $n$ is also similar, and $v$ is also comparable to the typical FWHM in the BLR of a few 1000~\kms, LoBALQs are likely lines of sight through the outskirts of BLR, before the total column builds up to $\NH> 10^{23}$~\cmmt, enough to completely obscure the inner parts of the AGN (e.g.\ Fig.~\ref{fig:synth_flux}, bottom panel) and turn it into a type 2 AGN. Obscuration will be significant if the gas is dusty, as observed for gas just outside the BLR \citep{suganuma_etal06}. If the BLR gas originates from a failed dusty disc wind \citep{czerny_hryniewicz11}, then the outskirts of the BLR will form a sheared radially driven failed wind, which would explain why only a blueshifted absorbing gas is generally seen. 

The BAL outflow produces CF$\sim 0.1$, compared to CF$\sim 0.3$ of the BLR gas \citep{korista_etal97, maiolino_elal01, ruff_etal12}. A line of sight below the outflow passes through the outskirts of the BLR, and produces LoBALQs. This outskirts region has CF$\sim 0.01$.  Some observational evidence for potential contribution of the BAL outflow to the BLR emission is discussed in Baskin et al. (in prep.).

\subsection{The absorber velocity structure}\label{sec:disc_vel_strcut}
What produces the velocity structure of the absorber? The hydrostatic solutions presented here do not address the velocity structure of the absorber. A time dependent hydrodynamic solution is required to address the expected profile (e.g.\ \citealt{proga_kallman04, giustini_proga12}). We note in passing that in the failed wind solution, the fast outflow may result from gas closer to the centre, which slows down as it accumulates more gas as it moves out. Eventually, the radial velocity drops close to zero, and the flow line falls back to the disc. The gas may be in a steady state, where our line of sight intersects the flow line of a continuous flow above the disc surface. Since the solution is non hydrostatic, only part of the radiation pressure compresses the gas, and the rest leads to an acceleration of the outflow. The lower compression will lead to a lower $n$ and a higher ionization. From the mass continuity equation, we also expect $n\times v$ to be conserved along flow lines in a steady state. The outflowing gas is expected to be more highly ionized at higher $v$. Indeed, in LoBALQs, the high-ionization lines extend to higher outflow velocities compared to the low-ionization lines. Also, in HiBALQs, the average outflow velocities extend to higher values, as the \HeII~$\lambda$1640 emission EW becomes weaker \citep{baskin_etal13}. The \HeII\ EW likely provides a measure of the hardness of the ionizing SED. A harder SED overionizes the illuminated gas to a larger column, where the gas is denser (Fig.~\ref{fig:NOdust_all_Nion_Sigma}), and thus eliminates the absorption down to smaller $v$  and higher $n$ along the flow line. A quantitative modelling of the observed magnitude of the effect (\citealt{baskin_etal13}; Baskin et al., in prep.) can be used to constrain the value of $n$ versus $v$, and allow one to measure the deviation from the hydrostatic solution.

\subsection{The expected response to changes in luminosity}
What is the expected response of the absorbing column to changes in the ionizing continuum? As discussed by \citet{chevallier_etal07} for warm absorbers, there are a few relevant timescales. First, the dynamical time $t_{\rm dyn}=D/c_s$, the time it will take the gas to adapt to a new steady state solution, where $c_s$ is the sound velocity, and $D$ is the width of the relevant layer. For a UV absorbing layer just outside the BLR, $D=\NH/n\simeq 10^{22}/10^9=10^{13}$~cm and $c_s\simeq 10^6$~cm~s$^{-1}$ which gives $t_{\rm dyn}\simeq 1$~yr. The ionization/recombination time scale is $t_{\rm ion}=1/(\alpha\times n)$, where $\alpha$ is the recombination rate, which gives $t_{\rm ion}\simeq 1$~hr. The light crossing time for a luminous $10^{46}$~\ergs\ AGN is $t_{\rm var}=r/c=10^{18}/3\times 10^{10}=1$~yr, but this is not relevant here, as we discuss changes only along the line of sight. For example, if the continuum changes significantly on a month timescale (typical for luminous quasars), then the gas ionization responds effectively instantly, while the gas density does not respond. As a result, $U$ at each point in the absorbing layer follows the luminosity changes. Since $\Nion$ is $\propto U$ (from Str\"{o}mgren depth consideration), the value of $\Nion$ of various ions will follow the change in luminosity. On $t> t_{\rm dyn}$ timescales, the hydrostatic solution holds, with the mean luminosity over the $t_{\rm dyn}$ timescale. For example, a transition from a constant low luminosity to a constant high luminosity on a month timescale, will produce a matching rise in $\Nion$, followed by a drop in $\Nion$ to the earlier values on a year timescale, despite the constant luminosity. This pattern of change in $\Nion$ is a unique signature of the RPC mechanism. The measurement of $t_{\rm dyn}$ through the response of the absorption to luminosity variations can be used to measure the distance of the absorber from the ionizing source, as $t_{\rm dyn}\propto D \propto r^2$.

\subsection{Dusty absorber}
There are three main differences between the dustless and the dusty absorber. First, the asymptotic value of \Nion\ is reached by $\NH\simeq 10^{20.5}-10^{21.5}$~\cmmt\ in a dusty absorber, compared to  $\NH\simeq 10^{22.5}$~\cmmt\ in a dustless absorber. Second, \Nion\ rises to its asymptotic value over a range of $\Delta\NH\approx 1$~dex in a dusty absorber, compared to a smaller range of $\Delta\NH\approx 0.1-0.2$~dex in a dustless absorber. Finally, the asymptotic value of \Nion\ is smaller by at least $\sim $0.5~dex for most ions in a dusty absorber. For example, the asymptotic \Nion\ of C$^{3+}$ and O$^{5+}$ is reduced by 1~dex to $\simeq10^{17}$~\cmmt, and for N$^{4+}$ it is reduced by 0.5~dex to $\simeq10^{16.5}$~\cmmt. The reduction in the asymptotic \Nion\ of Si$^{3+}$ is significantly larger, by a factor of 2~dex to $\simeq10^{14.7}$~\cmmt, since Si is assumed to be heavily depleted into grains.

While the predicted maximal \Nion\ for the \CIV\ BAL is consistent with observations (see below),  the predicted  \Nion\ for Si$^{3+}$ in a dusty absorber is too low to produce the observed typical \SiIV\ absorption. Approximately 50 per cent of \CIV\ BALQs also present \SiIV\ BAL absorption (e.g.\ \citealt{gibson_etal09}), which is marginally saturated \citep{baskin_etal13}. This \SiIV\ absorption requires $\Nion\approx10^{16}$~\cmmt, which is larger by $\ga$1~dex than the predicted maximal \Nion\ of Si$^{3+}$ in a dusty absorber. This apparent inconsistency may result from the assumed dust composition. The \cl\ model assumes Galactic ISM dust (Section~\ref{sec:model_numerical}), while the reddening observed in BALQs is remarkably well fit by SMC dust \citep{sprayberry_foltz92, baskin_etal13}, where some studies suggest there is almost no Si depletion (\citealt{welty_etal01}; cf.\ \citealt*{sofia_etal06, li_etal06}).

What is the expected reddening associated with \Nion? The maximal \NHtot\ allowed to avoid significant Lyman edge absorption, which is not seen in HiBALQs \citep{baskin_etal13} is $\simeq10^{21.5}$, $10^{21}$ and $10^{20.5}$~\cmmt\ for the $Z/Z_{\sun}=0.5$, 1 and 5 model, respectively (Fig.~\ref{fig:dust_all_Nion_Sigma}). These values of \NHtot\ correspond to $E(B-V)\approx0.2$~mag, assuming that the $E(B-V)/\NHtot$ ratio of Galactic ISM (\citealt{draine_11}, equation 21.6) scales linearly with $Z$. The $E(B-V)/\NHtot$ ratio is reported to be a factor of 2--10 lower for SMC dust than in the Galactic ISM \citep{sofia_etal06}, which implies $E(B-V)\approx0.02-0.1$~mag. This range of possible values of $E(B-V)$ is  consistent with the upper limit of 0.04~mag estimated for HiBALQs assuming SMC dust (\citealt{baskin_etal13}; \citealt{gibson_etal09} report 0.02~mag).

A major difficulty with the dusty absorber interpretation is the relatively small column of $\NHtot\simeq 10^{21}$~\cmmt\ allowed in HiBALQs, to avoid significant Lyman edge absorption. This column stands in contrast with the two indicators, X-ray absorption and the \PV\ line, which suggest $\NHtot\sim 10^{22}$~\cmmt, as discussed below.

\subsection{Comparison with observations}\label{sec:compare_obs}

\subsubsection{The X-ray column}
As noted above (Section~\ref{sec:intro}), although BALQs tend to be X-ray weak, their weakness does not appear to be generally a result of the large column ($>10^{23}$~\cmmt) required to extinct the observed emission below a few keV. X-ray spectroscopy of BALQs generally reveals absorption, but the inferred columns are typically in the range of 1--3$\times 10^{22}$~\cmmt\ which provide only minor absorption above 1~keV \citep*{gallagher_etal06, giustini_etal08, streblyanska_etal10, hamann_etal13, morabito_etal14}. RPC in HiBALQs implies a total column of $\simeq 3.5\times 10^{22}$~\cmmt, consistent with the observational results. However, the effective X-ray absorbing column is likely only $\simeq 1-2\times 10^{22}$~\cmmt, as the surface layer of RPC gas is too highly ionized to produce absorption. The X-ray spectroscopy available for BALQs is generally not sufficient to determine the ionization level of the absorber, and the measured $\NHtot$ are commonly made assuming a neutral absorber. It will thus be interesting to try and infer the implied $\NHtot$ in BALQs based on RPC photoionization modelling fits to the available data. Furthermore, it will be interesting to obtain high quality X-ray spectroscopy for BALQs, so one can test if the various spectral features predicted by RPC (Fig.~\ref{fig:synth_flux}) are observed. 

High quality X-ray spectroscopy is available for a handful of nearby AGN. These AGN are not BALQs, but do reveal some UV absorption lines with typical line widths of a few hundred \kms. The high quality X-ray spectra allow one to derive the absorption measure distribution over a wide ionization range, and the results are remarkably well fit by the RPC models, as further discussed in paper III. It will be interesting to extend these studies to BALQs.

\subsubsection{The UV ionic columns}
Below we compare the asymptotic \Nion\ of different ions with the observed columns. A knowledge of \NHtot\ will allow us to get the specific predicted \Nion, but such information is generally unavailable. In principle, a measurement of the H$^0$ column can be used to estimate \NHtot\ (e.g.\ Fig.~\ref{fig:NOdust_Nion_Sigma}). However, such a measurement requires a high-S/N spectral coverage down to the Lyman limit (e.g.\ \citealt{arav_etal01b}), which is usually unavailable.

The dustless absorber is qualitatively consistent with observations. The high-ionization BALs of abundant elements (\SVI, \OVI, \NV\ and \CIV) appear to be saturated both in the average absorption spectrum of a large sample of BALQs \citep{baskin_etal13}, and in the absorption spectrum of individual objects (e.g.\ \citealt{arav_etal01a, arav_etal01b, moe_etal09, borguet_etal12}). This implies $\Nion\ga10^{16}$~\cmmt, which is similar to the asymptotic \Nion\ of S$^{5+}$, and is well below the asymptotic \Nion\ of $\simeq10^{17}$~\cmmt\ for N$^{4+}$, and $\simeq10^{18}$~\cmmt\ for O$^{5+}$ and C$^{3+}$ (Fig.~\ref{fig:NOdust_Nion_Z_and_alpha}). The lower-ionization \SiIV\ BAL is typically marginally saturated \citep{baskin_etal13}, which implies $\Nion\la10^{16}$~\cmmt. The measured \SiIV\ columns for PG~0946+301, QSO 2359-1241, SDSS J0838+2955 and SDSS J1512+1119 are $>15.1$, $>15.4$, $16.4$ and $>14.71$~\cmmt, respectively \citep{arav_etal01b, arav_etal01a, moe_etal09, borguet_etal12}. These values are consistent with the asymptotic \Nion\ of $\simeq10^{17}$~\cmmt\ for Si$^{3+}$ (Fig.~\ref{fig:NOdust_Nion_Z_and_alpha}). Another general property of BALQs is that strong LoBALs (e.g.\ \MgII\ and \AlIII) are always accompanied by strong HiBALs. This phenomenon is readily explained by RPC, since an RPC slab which produces absorption in low-ionization lines \emph{inevitably} also produces absorption in high-ionization lines, by layers closer to the illuminated face (Figs~\ref{fig:NOdust_Nion_Sigma} and \ref{fig:dust_Nion_Sigma}). This is in contrast with uniform-density models, which typically require two distinct, low- and high-$U$, slabs to reproduce both types of absorption lines.

A more detailed comparison between RPC models and observations can be carried out for two objects. For PG~0946+301, \citet{arav_etal01b} derived  $15.4<\log\Nion({\rm H}^0)<17$. The reported lower limits on $\log\Nion$ for C$^{3+}$, N$^{4+}$, O$^{5+}$, P$^{4+}$, Si$^{3+}$ and S$^{5+}$ are 16.1, 16.2, 16.6, 15.0, 15.1 and 15.8, respectively ($\log\Nion<14.7$ for C$^+$). A dustless RPC slab with $\NHtot\la3.8\times10^{22}$~\cmmt\ (Fig.~\ref{fig:NOdust_Nion_Sigma}) has \Nion\ values that are consistent with the measured lower limits for all ions, except P$^{4+}$, where the predicted maximal log column is $<14.5$ (Fig.~\ref{fig:NOdust_Nion_Z_and_alpha}).\footnote{In order to reproduce the observed P$^{4+}$ column by a uniform-density photoionization model, \citet{arav_etal01b} require an enhanced P abundance by a factor of $\sim 10$ relative to other metals compared to the solar abundance.} For SDSS J0838+2955, \citet{moe_etal09} use the non-saturated \AlIII\ and \SiIV\ doublets to evaluate the absorber CF, and report $\log\Nion$ of 14.9, 16.4, 13.9, 14.6 and 16.1 for C$^+$, C$^{3+}$, Mg$^+$, Al$^{++}$ and Si$^{3+}$, respectively. For log \Nion(Si$^{3+}$)$\simeq16$, all dustless RPC models predict log\Nion(C$^{3+})\simeq 17.5$ (Fig.~\ref{fig:NOdust_all_Nion_Sigma}) which is 1~dex larger than the measured C$^{3+}$ column. This discrepancy may be caused by deviations of the observed absorber from hydrostatic equilibrium assumed by the RPC models.  A more detailed fit to observations is beyond the scope of this paper.

A detection of P$^{4+}$ in BALQs based on the \PV\ $\lambda\lambda$1118, 1128 lines was suggested by \citet{hamann_98}. The estimated ionic column of $\sim 10^{15}$~\cmmt\ implies a total column of $\NHtot \sim 10^{22}$~\cmmt\ given the low abundance of P. This column is consistent with the RPC results. The data quality available in \citet{hamann_98} is rather low, but additional evidence may be forthcoming \citep{hamann_etal13b}. A high resolution and high S/N spectrum with \PV\ $\lambda\lambda$ 1118, 1128 absorption is presented by \citet{borguet_etal12}, which yields log$\Nion=14.6$, remarkably close to the RPC prediction (Fig.~\ref{fig:NOdust_Nion_Z_and_alpha}). However, the absorbing system has a FWHM of only $\sim 300$~\kms\ and is clearly not a BAL system, and likely originates much further out, where as noted above, the RPC solution is also expected to apply.

\subsection{Additional physical implications}
The RPC mechanism implies a total absorber column of $\NHtot\simeq 3.5-4\times 10^{22}$~\cmmt. This mechanism is present regardless of the mechanism which drives the outflow. If the outflow is radiation pressure driven, then there is a minimal value of \LLedd\ required to drive the outflow. If the absorber is dustless, then the mean $a/a_{\rm es}\approx 10$ (Fig.~\ref{fig:slab_acc}, left panel) implies one needs $\LLedd>0.1$ to produce an outflow. Such an outflow will have a terminal velocity comparable to the local Keplerian velocity, and $\LLedd\sim 1$ is required to produce an outflow a factor of a few faster, as commonly observed.

If the absorber is dusty, then it has $\NHtot\simeq 10^{21}$~\cmmt\ to avoid a Lyman edge (Fig.~\ref{fig:dust_Nion_Sigma}, left panel), and a mean $a/a_{\rm es}\approx 500$ (Fig.~\ref{fig:slab_acc}, right panel). In this case, $\LLedd>0.002$ is enough to start an outflow, and a terminal velocity well above Keplerian is possible already at moderate \LLedd\ values. However, such an outflow will not produce the commonly observed X-ray absorption in HiBALQs, and the \PV\ absorption which may also be common \citep{hamann_etal13b}.

The handful of BALQs in the PG quasar sample do tend to cluster near $\LLedd\sim 1$ (\citealt{laor_brandt02}, fig.~9 there), which also supports the dustless outflow interpretation. Clearly, if this distribution holds in a large sample, this will provide a strong support that the outflow in BALQs is radiatively driven.

How stable is the outflow? The interaction of radiation and matter is often expected to lead to radiation Rayleigh-Taylor instability, and the development of such systems beyond the instability timescale calls for numerical simulations (e.g.\ \citealt*{krumholz_matzner09, krumholz_thompson13, jiang_etal13}). A recent hydrodynamical simulation of a gas cloud exposed to AGN radiation is presented by \citet*{namekata_etal14}. The simulation shows the formation of an effectively RPC slab structure (i.e.\ an exponential density rise with depth) from the initial conditions of a uniform-density spherical cloud (e.g.\ figs~13 and 18 there). In the presence of a large supporting column ($>10^{24}$~\cmmt) such a cloud may be effectively stable. However, this configuration is relevant for the BLR gas (paper II). In a lower column outflowing gas, the structure is likely transient, possibly consistent with the observed variability of the absorption profiles in BALQs. 

As noted in Section~\ref{sec:model}, confinement on the outward side of the outflowing sheet can be provided by the ram pressure of the ambient medium. What is the associated heating of the gas, compared to the heating by the ionizing radiation incident from the inward direction?  The energy flux of the ionizing radiation is $P_{\rm rad}c$. The energy flux from the confining gas on the outward side is $P_{\rm ram}v$, where $P_{\rm ram}$ is the ram pressure and $v$ is the outflow velocity. Since $P_{\rm ram}=P_{\rm gas}$ to provide confinement, and $P_{\rm gas}=P_{\rm rad}$ from RPC, the heating ratio $P_{\rm ram}v/P_{\rm rad}c$ is just $v/c$, or a few per cent. The heating by the ram pressure is therefore negligible. However, if the ambient medium is dense enough, the mass build up may be large enough to slow down the outflow and turn it into a failed wind.

\section{Conclusions}\label{sec:conclusions}
BAL outflows can reach velocities well above 10,000~\kms, and thus most likely originate on sub-kpc scales, and possibly on sub-pc scales, i.e.\ just outside the BLR. A likely origin is a wind from the accretion disc. At such a small distance from the ionizing source, the wind will be highly ionized, unless the ionizing radiation is significantly absorbed before reaching the wind, or if the wind material is highly clumped. Recent observations have clearly ruled out the absorption scenario \citep{hamann_etal13}, implying that the wind material must fill $<10^{-3}$ of the volume along our line of sight to prevent overionization. Here we point out that RPC provides a natural mechanism for the low observed filling factor. The ionizing radiation compresses and confines the gas along the line of sight, producing gas dense enough to prevent overionization, but ionized enough to produce the observed UV absorption lines. 

Based on photoionization calculations using \cl\ we find the following.
\begin{enumerate}
\item In contrast with a uniform-density absorber, in RPC absorber there is a large gradient in ionization states along the line of sight. Most of the UV absorption observed in HiBALQs occurs in a layer at $\NH=3-4\times 10^{22}$~cm$^{-2}$. The layer at $\NH<3\times 10^{22}$~cm$^{-2}$ is highly ionized and produces only X-ray absorption (paper III), while the layer at $\NH>4\times 10^{22}$~cm$^{-2}$ produces the low-ionization lines observed in LoBALQs. 

\item The structure of the absorbing layer is independent of distance from the AGN. It is predicted to produce highly saturated absorption lines from ions of C, N, O and Ne, for the typical observed BAL velocity dispersion of a few 1000~\kms. The less abundant elements of Mg, Si, S and Fe can produce optically thick broad lines. The lower abundance elements of Na, Al, Ar and Ca can produce detectable absorption, while even lower abundance elements such as P, Cl and K, can be detectable if the absorption lines are narrow.

\item The shape of the ionizing SED has a negligible effect on the predicted \Nion\ of the various lines. The effect of absorber metallicity on the strength of the low-ionization lines is roughly linear. The strength of the higher-ionization lines, excluding N, is only weakly dependent on the metallicity.

\item If the RPC absorber is dusty, the ionized layer has a column of only $\sim 10^{21}$~\cmmt. However, the observed X-ray absorption and the detection of \PV\ absorption, imply a column of $\sim 10^{22}$~\cmmt, which is consistent with the RPC prediction for dustless gas.

\item It is not clear why $\sim 90$ per cent of BALQs avoid $\NH>4\times 10^{22}$~\cmmt. A hint may be provided by the sharp change in the gas opacity at the He$^+$ and the H ionization fronts, which occur near this column. These sharp changes may lead to a dynamical instability in a radiation pressure driven flow, which may affect the survival of such absorbers.

\item The RPC derived column and ionization structure of the absorbing gas imply that BALQs should have $\LLedd>0.1$, if the outflow is radiatively driven.
\end{enumerate}
The results presented above are based on the hydrostatic RPC simulation. The observed high velocity dispersion of the absorption lines in BALQs implies that at least part of the radiation pressure is used towards accelerating the outflow, and only a part can be used to compress the gas.

Clearly, a more realistic modelling requires a time dependent solution which follows the dynamics of low-density extended gas distribution, likely injected from the disc surface, as it becomes exposed to the ionizing radiation. The gas will be compressed and accelerated, and together with photoionization calculations, such models can predict the detailed absorption profiles of the outflows. Even if RPC is only a 10 per cent effect, it can still compress the gas enough to avoid overionization. The set of predicted \Nion\ values can be used to test observationally the validity of the RPC solution for BALQs.

\section*{Acknowledgements}
We thank F.\ Hamann for many valuable comments, the anonymous referee for comments and suggestions, and G.\ Ferland for developing \cl\ and making it publicly available.
AL acknowledges support through grant 2017620 from the Helen and Robert Asher Fund at the Technion.
This research has made use of NASA's Astrophysics Data System Bibliographic Services.

\bsp
\label{lastpage}
\end{document}